\documentstyle[12pt,epsf]{article}

\def\thefootnote{\fnsymbol{footnote}}

\newcommand{\eq}{\begin{equation}}
\newcommand{\en}{\end{equation}}
\newcommand{\be}{\begin{equation}}
\newcommand{\ee}{\end{equation}}
\newcommand{\eqa}{\begin{eqnarray}}
\newcommand{\ena}{\end{eqnarray}}
\newcommand{\ba}{\begin{eqnarray}}
\newcommand{\ea}{\end{eqnarray}}

\newcommand{\Br}{\langle}
\newcommand{\kt}{\rangle}

\newcommand{\um}{\frac12}

\newcommand{\ZZ}{\hbox{{\rm Z{\hbox to 3pt{\hss\rm Z}}}}}
\def\de{\partial}

\newcommand{\simlt}
{\mathrel{\raisebox{-.3em}{$\stackrel{\displaystyle <}{\sim}$}}}

\newcommand{\NP}[1]{Nucl.\ Phys.\ {\bf #1}}

\begin{document}
\begin{titlepage}
\vskip0.5cm
\begin{flushright}
DFTT 33/02\\
DESY 02-180\\
\end{flushright}
\vskip0.5cm
\begin{center}
{\Large\bf  String  effects in the}
\vskip 0.3cm
{\Large\bf  3d gauge Ising model}
\end{center}
\vskip 1.3cm
\centerline{
M. Caselle$^a$, M. Hasenbusch$^{b}$
 and M. Panero$^a$}
 \vskip 1.0cm
 \centerline{\sl  $^a$ Dipartimento di Fisica
 Teorica dell'Universit\`a di Torino and I.N.F.N.,}
 \centerline{\sl via P.Giuria 1, I-10125 Torino, Italy}
 \centerline{\sl
e--mail: \hskip 1cm
 caselle@to.infn.it \hskip 1cm
 panero@to.infn.it}
 \vskip .4 cm
 \centerline{\sl  $^b$ NIC/DESY Zeuthen, Platanenallee 6, D-15738 Zeuthen,
 Germany}
 \centerline{\sl
e--mail: \hskip 1cm
 Martin.Hasenbusch@desy.de}
 \vskip 1.cm

\begin{abstract}
We compare the predictions of the effective string description of confinement with
a set of Montecarlo data for the 3d gauge Ising model at finite temperature. 
Thanks to a new algorithm
which makes use of
 the dual symmetry of the model we can reach very high precisions even for
 large quark-antiquark distances. We are thus able to explore the large $R$
regime of the effective string. We find that for large enough distances and low
enough temperature the data are well described by a pure bosonic string. As the
temperature increases higher order corrections become important and cannot be
neglected even at large distances. These higher order corrections seem to be
well described by the 
Nambu-Goto action truncated at the first perturbative order.

\end{abstract}
\end{titlepage}

\setcounter{footnote}{0}
\def\thefootnote{\arabic{footnote}}

\section{Introduction}
In these last years  
a lot of efforts have been devoted to extract  the interquark potential from 
lattice gauge theories (LGT's) looking at the expectation values of Wilson 
loops or Po\-lya\-kov loop correlators in Montecarlo 
simulations. Besides the important goal of obtaining reliable  
values of physical observables like the string tension,  
these simulations also allow to study the physical
nature of the potential. In particular, a very interesting issue is the so
called ``string picture'' of the interquark potential: 
quark and antiquark linked 
together by a thin fluctuating  flux tube~\cite{conj}. 

\vskip.3cm

The standard approach to study this problem is to look at the finite size 
effects due to quantum string fluctuations, which, in finite geometries, 
give measurable contributions to the interquark potential. 
This approach traces back to the seminal work of L\"uscher, Symanzik and 
Weisz~\cite{lsw} and has interesting connections with the conformal 
field theory (CFT) approach of two-dimensional models developed in the 
eighties~\cite{bpz,cft}. 
\vskip.3 cm

The major problem in trying to use these finite size corrections to obtain
information on the underlying effective string is that 
very high precision estimates of the
interquark potential are
 needed.
Such a precision is very hard to reach with standard algorithms, in
particular if one is interested in the large distance regime where the effective
string should show up.

This led us in the past years to concentrate on the simplest non-trivial LGT,
namely the 3d gauge Ising model for which, thanks to the dual 
transformation, new
and very powerful algorithms can be constructed and very 
high precisions can be reached within a
reasonable amount of CPU-time even for
large distance interquark potentials.
Following this line we found convincing evidences
for the existence of a bosonic type effective string theory in the 3d gauge
Ising model both in the case of the Wilson loop geometry~\cite{wloops}
 (fixed boundary
conditions (b.c.) in both directions) and of the
interface geometry~\cite{cfghpv} (periodic bc in both directions). This paper deals with
the third remaining case, namely that of the Polyakov loop correlators, which
corresponds to a mixed geometry (fixed b.c. in one direction and 
periodic b.c. in the other direction). A preliminary account of the present
study
recently appeared in~\cite{cpp02}. Here we complete the analysis using a new
algorithm which fully exploits the power of dual transformations, leading to
a gain of more than one order of magnitude in
precision with respect to~\cite{cpp02}. Thanks to this higher resolution we are
now able 
 to explore in greater detail the fine structure
 and the higher order corrections of the underlying effective string
theory. This is the main goal of the present paper.

\vskip .3cm
While studying the Ising model allows a very careful control of all
possible sources of systematic errors and a very precise study of the
fine details of the underlying effective string, 
it remains an open problem to see if
the results obtained  are particular features of the gauge 
Ising model only or have a more
general validity and can be extended also to non-abelian LGT which are 
more interesting from a physical point of view. 
\vskip.3 cm

Last year, an important progress was made in this direction,
thanks to a new, powerful, 
algorithm proposed by L\"uscher and Weisz in~\cite{lw01}. 
With such an algorithm an
exponential reduction of statistical errors of Polyakov loop correlators 
can be obtained with no need of dual transformations. Hence it can be used
for any non-abelian LGT, thus allowing to study the possible existence of string
corrections in a much wider set of models. In particular it was recently 
used by the same authors in~\cite{lw02} to study the
SU(3) theory both in (2+1) and (3+1) dimensions. In both cases they found again
a good agreement with the predictions of the free bosonic effective string
theory.

Thanks to this relevant progress it is now possible to address the important
issue of the string universality, i.e.
to compare the properties of the effective strings underlying 
different LGT's. It is by now clear that 
for large enough distances (and low enough temperatures
in the case of Polyakov loop correlators) one always finds
 the same asymptotic theory,
i.e. the free bosonic effective theory originally studied in~\cite{lsw}.
However for shorter distances and/or higher temperatures, terms of higher order
(typically self-interaction terms or boundary-type contributions)
which are present in the string action start to give measurable
corrections and can be detected and studied. 

In this respect Polyakov loop correlators turn out to be a perfect tool to study
these effects, since as the temperature increases these higher order
corrections become rather large even for large interquark distances, i.e. in a
regime in which other possible sources of corrections (say, for instance,
the perturbative one gluon exchange contribution in SU(3) ) 
are under control or negligible.
 
Thanks to this fact and to the relevant precision of our simulations we are able
to precisely observe the deviations with respect to the free bosonic string
predictions, and we can see that, as expected, they increase in magnitude as the
temperature increases. We shall also show that these corrections are well
described (with some cautionary remarks discussed in sect. 4 and 5 below)
by a Nambu-Goto type string action, truncated at the first
perturbative order.

\vskip.3 cm
This paper is organized as follows. In sect. 2 we shall discuss some general
results concerning the effective
 string description of the interquark potential. 
 We shall introduce both the bosonic string and the self-interaction terms
 induced by the Nambu-Goto string. We shall then discuss the corresponding 
 finite
 size corrections. We made an effort to make this section as self-contained as
 possible, so as to allow the reader to follow all the steps of the derivation.
  In sect. 3 we shall give a few general
information on the 3d gauge Ising model, on the algorithm that we used to
simulate the model, and we shall also describe in some detail the 
simulations that we performed. In sect. 4 we shall discuss our results. Finally
sect. 5 will be devoted to some concluding remarks and to a comparison with the
analogous results obtained in SU(3) in~\cite{lw02}.

\section{Theory}
\label{s2}

As mentioned in the introduction, 
Polyakov loops naturally arise if one studies finite temperature LGT. 
Thus we shall begin this section with a brief summary of known results on
finite temperature LGT (sect. \ref{s2.1}). This will also allow us to fix 
notations and conventions. Next we shall address the issue of finite size
corrections, following two complementary paths. First we shall discuss 
them in full generality, without resorting to any specific string model, but
simply exploiting the quantum field theory implications of the roughening
transition (sect. \ref{s2.2}), using some general results of Conformal Field
Theory (sect. \ref{s2.3}) and the exact solution of the CFT of a free boson
(sect. \ref{s2.4}). This approach is very powerful but it allows no insight in
possible higher order corrections due to the self-interaction of the string. To
this end a precise choice of the effective string model (i.e. the precise form
of world-sheet Lagrangian of the underlying string) is needed. We shall address
this point in sect. \ref{s2.5}, studying the simplest possible string (the
natural 
 generalization  
of the free bosonic CFT) i.e. the Nambu-Goto string.
Next in sect. \ref{s2.6} we shall outline the implications of this choice for
our understanding of the deconfinement transition. Finally we close this
theoretical
introduction by addressing the important issue of the range of validity 
of the effective string picture (sect. \ref{s2.7}).

\subsection{Finite temperature gauge theories: general setting and notations}
\label{s2.1}
The partition function of a gauge theory in $d$ spacetime dimensions
 with gauge group $G$ 
regularized on a lattice is
\eq
Z=\int\prod \mbox{d}U_l(\vec x,t) 
\exp\{-\beta\sum_p\, \mbox{Re}\, \mbox{Tr}(1-U_p)\}~~~,
\label{eq1}
\en
where $U_l(\vec x,t)\in G$ is the link variable
at the site  $(\vec x,t)=(x_1,..,x_{d-1},t)$ in the direction $l$ and
$U_p$ is the product of the 
links around the plaquette $p$.

Let us call $N_t$ ($N_s$) the lattice size in the
time (space) direction (we assume for simplicity $N_s$ to be 
the same for all the space directions).
Lattice simulations with non-zero temperature
are obtained by imposing periodic boundary conditions in the time direction.
 A $(N_s)^{d-1}N_t$ lattice can then be interpreted as representing 
a system of finite volume $V=(N_sa)^{d-1}$ at a finite temperature
$T=1/L=1/N_ta$ where $a$ is the lattice spacing. To simplify notations we shall
fix from now on the lattice spacing to be 1 and neglect it in the following.
\vskip.3 cm

The order parameter of the finite temperature deconfinement transition
 is the Polyakov loop, $i.e.$  the trace of the ordered product of all
 time links with the same space coordinates; this loop is
 closed owing  to the periodic boundary conditions in the time direction:
\eq
P(\vec x)=\mbox{Tr}\prod_{z=1}^{N_t}U_t(\vec x,z)~~~.
\en
The vacuum expectation value of the Polyakov loop is zero in the
confining phase and acquires a non-zero expectation value in the 
deconfined phase. The value $\beta_c(T)$ of this deconfinement 
transition is a function of the temperature, and defines a new 
physical observable $T_c$. 
The inverse of this function gives for each value of $\beta$ 
the lattice size in the time direction (which we shall call in
the following $N_{t,c}(\beta)$) at which the model undergoes the deconfinement
transition.
\vskip.3 cm

The interquark potential can be extracted by looking at the correlations  
of Polyakov loops in the confined phase.
The correlation of two loops $P(x)$  at
a distance $R$ and at a temperature $T=1/L=1/N_t$ is given by 

\eq
G(R)\equiv 
\langle P(x)P^\dagger(x+R) \rangle \equiv {\rm e}^{-F(R,L)}~~~,
\label{polya}
\en
\noindent
where  the free energy $F(R,L)$ is expected to be described, as a 
first approximation,  by the so called ``area law'':

\eq
F(R,L)\sim F_{cl}(R,L)=\sigma L R + k(L)
~~~,\label{area}
\en
where $\sigma$ denotes the string tension\footnote{In 
the following, when needed, we shall also explicitly
write the dependence of the string tension on the finite temperature $T$ and the
coupling $\beta$ as $\sigma(T)$ or $\sigma(T,\beta)$ depending on the case} 
and $k(L)$ is a non-universal
 constant depending only on $L$. The meaning of the index $cl$ refers to the
 fact that (as we shall discuss below) this should be considered as
 a ``classical'' result, which neglects quantum fluctuations.

In the following we shall mainly study the combination
\eq
Q_0(R,L)\equiv F(R+1,L)-F(R,L)\equiv\log\left(\frac{G(R)}{G(R+1)}\right)
\en
in which the non-universal constant cancels out.

The observable (\ref{polya}) is similar to the expectation value of an ordinary
 Wilson loop except for the boundary conditions, which are  in this case 
fixed in the space directions and periodic in the time direction. The
 resulting geometry is that of a cylinder, which is topologically 
different from the rectangular geometry of the Wilson loop.

\subsection{The roughening transition and the effective string}
\label{s2.2}

Eq.(\ref{area}) correctly describes the Polyakov loop correlators only in the
strong coupling phase. As it is well known
the confining regime of a generic lattice gauge theory 
 consists in general of two phases: the 
strong coupling phase and the rough
phase. These two phases are separated by the roughening transition where
the  strong coupling expansion for the Polyakov
 loop correlator (as well as that for the Wilson loop or the interface)
ceases to converge~\cite{rough,lsw}.
 These two phases are related to
two different behaviors of the quantum fluctuations of the 
flux tube around its
equilibrium position~\cite{lsw}. 
In the strong coupling phase, these fluctuations are
massive, while in the rough phase they become massless 
and hence survive in the
continuum limit. The inverse of the mass scale of these
fluctuations
\footnote{Notice that this scale
 is completely different
from the glueball mass scale.}
can be considered as a new 
correlation length of the model. It is exactly this new correlation length
 which goes to infinity at the roughening point and
induces the singular behavior of the strong coupling expansion. 
\vskip 0.3cm
In the rough phase the 
 flux--tube fluctuations can be described by a suitable
two-dimensional massless quantum field theory,
where the fields describe the
transverse displacements of the flux tube. 
 The common lore is that this QFT should be the effective 
low energy description of some fundamental string theory (this is the reason for
which this QFT is often called ``effective string theory'' and 
 the finite size contributions it induces are usually named
``string corrections'' )\footnote{Notice however that the existence of
such an underlying fundamental string theory is not a mandatory requirement
 to justify the results that we
shall discuss below. Any alternative mechanism (see for instance ~\cite{gt01})
 which could induce a fluctuating
flux tube description for the interquark potential works equivalently well.}.
It is
expected to be very complicated
 and to contain in general non-renormalizable interaction terms
\cite{lsw}.
However, exactly because these interactions are non-renormalizable, their
contribution is expected to be
 negligible in the infrared limit (namely for large quark
separation)~\cite{olesen}.
In this infrared limit the QFT becomes a conformal invariant
field theory (CFT)~\cite{cft}.

From the general theory of CFT's  we immediately see that there are
 two important signatures which, if detected, could validate the whole
 picture,
 and which could be in principle observed in numerical
 simulations.
\begin{description}
\item{(1)} The massless quantum fluctuations delocalize the flux tube
 which acquires a nonzero width, which diverges logarithmically as the
 interquark distance increases~\cite{lmw, width}.
\item{(2)} These quantum fluctuations
 give a non-zero contribution to the interquark potential,
which is related to the partition function of the above 2d QFT.
Hence if the 2d QFT is simple enough to be exactly solvable (and this is in
general the case for the CFT in the infrared limit) also these
contributions can be evaluated exactly. They show up as finite size
corrections to the interquark potential.
\end{description}

 It is this last signature which is the
 best suited to be studied by numerical method and which we shall address in the
 following section.

\vskip 0.5cm
\subsection{ Finite Size Effects: general discussion}
\label{s2.3}
As mentioned above, the pure area law
 is  inadequate to describe the Polyakov
loop correlator in the rough phase and
 must be  multiplied by the 
partition function of the 2d QFT describing  the quantum
fluctuations of the flux tube which
 in the infrared limit becomes a 2d CFT. Let us call
  $Z_{q}(R,L)$ the  partition function  of such a CFT on the cylinder (the open
  ends of the cylinder being the two Polyakov loops). Then
eq.~(\ref{area}) in the rough phase becomes:

\eq
\langle P(x)P^\dagger(x+R) \rangle = {\rm e}^{-F_{cl}(R,L)}
Z_{q}(R,L)~~~~.
\label{polya2}
\en

Defining the  free energy of quantum fluctuations as
$$F_q(R,L)=-\log Z_q(R,L)~~~~,$$ 
we find for the free energy
\eq
F(R,L)\sim F_{cl}(R,L)+ F_q(R,L)=\sigma L R + k(L)
- \log{Z_q(R,L)}
~~~.\label{a+q}
\en

By using standard CFT's~\cite{cft} methods
 we can study the behavior of 
$F_q(R,L)$ as a function of $R$ and $L$ 
in a general way. Indeed any two dimensional 
CFT is completely described once the conformal anomaly $c$, 
the operator content $h_i$ and the operator product algebra (or the 
fusion algebra which equivalently encodes all the fusing properties
of the CFT) are given. Then it is easy to show that $F_q(R,L)$ only 
depends on the adimensional ratio\footnote{the factor of 2 in the definition of
$z$ is a consequence of the asymmetry in the boundary conditions.} $z=2R/L$. It is possible to 
give asymptotic expressions for $F_q(R,L)$ in the $z\gg 1$ and 
$z\ll 1$ regimes: 

$z\gg 1$~\cite{bcna}:
\eq
F_q(R,L)\simeq-\tilde c \frac{\pi R}{6 L}~~~,
\label{zbig}
\en
where  
$\tilde c=c-24h_{min}$ is the effective conformal anomaly \cite{ISZ}; 
 $h_{min}$ is the lowest conformal weight of the physical states  
propagating along the cylinder. In the case of unitary CFT's $h_{min}=0$ 
(unless special boundary conditions are chosen) 
and $\tilde c$ coincides with the conformal anomaly $c$;

$z\ll 1$~\cite{cardy}:
\eq
F_q(R,L)\simeq-\hat c \frac{\pi L}{24 R}~~~,
\label{zsmall}
\en
where $\hat c=c-24h_{\alpha,\beta}$ and 
$h_{\alpha,\beta}$ is the lowest conformal weight compatible with the  
boundary conditions  $\alpha$ and $\beta$ at the two open  ends of the 
cylinder. In the case of an unitary CFT and fixed b.c. we have again
$h_{\alpha,\beta}=0$.
             
If the CFT is exactly solvable, namely if the whole operator content is 
known, one can explicitly write the free energy for all values of $z$, 
which smoothly interpolates between the two asymptotic behaviors.
\vskip .3cm

An important role in this construction is played by the modular 
transformations. All the partition functions can be written as 
power expansions in $q=\exp({2\pi i \tau})$, with $\tau=iz$
for Polyakov loop correlations
(notice that $\tau=i\frac{R}{L}$ if one studies Wilson loops).
Modular transformations allow to extend these expansions in the whole 
$\tau$ plane. In particular we shall be interested, in the following, in
the $\tau \to -1/\tau$ transformation.
\vskip .3cm

In the Wilson loop case, this 
transformation is a symmetry, because it exchanges $R$ and $L$. 
We can use this symmetry by choosing 
for instance $L\geq R$, and $\tau=iL/R$. With this choice
 $L$ plays the role of a time-like extent
 and  the interquark
potential $V(R)$ we want to extract from the data is defined in the
limit: $V(R)=\lim_{L\to\infty}F(R,L)/L$.
A similar symmetric situation occurs if one studies the behavior of the
interface tension (see for instance~\cite{cfghpv}).
\vskip .3cm

In the Polyakov loop case, the situation is completely different: $L$ and $R$ have
a  different meaning and the modular transformation
$\tau\to-1/\tau$ allows us to move from the region in which $2R>L$ to 
that in which $2R<L$. What is new is that, due to the modular
transformation, in these two regions the string corrections have, as we 
have seen above, different functional forms.
While in the region in which $2R<L$ the dominant 
contribution is, like in the Wilson loop case, of the type $1/R$,
in the region 
in which $2R>L$ the dominant contribution is proportional to $R$, 
and acts as a finite size correction of the string tension. This 
behavior will play a major role in the following.

\subsection{The simplest case: the free bosonic string}
\label{s2.4}

The simplest possible choice for the CFT which should describe the effective
string in the infrared limit is to assume that the $d-2$ fields which describe
the transverse displacement of the flux tube are $d-2$ free non-interacting
bosons.
We shall denote in the following this approximation of $F_q$ with the
notation $F_q^1$ and the corresponding partition function as $Z_1$\footnote{
The rationale behind this choice is that we think of $F_q^1$ as the first term 
in the expansion of $F_q$ in powers of $(\sigma RL)^{-1}$. 
We shall address below the second
term of this expansion which we shall denote as $F_q^{NLO}$.}.

 With abuse of language this choice is usually referred to as the ``bosonic
string'' model. Notice however that this model, being only 
 an effective long range description, could well be
related to a wide class of wildly interacting (and not necessarily bosonic)
 string theories. Besides being the simplest choice this model is 
 very important for at least three reasons:
\begin{description}
\item{1]} In the framework of the interface physics this QFT is known as
\emph{``capillary wave model''} and has received in the past years impressive confirmations
in a set of studies of different models belonging to the Ising universality
class (see for instance ref.~\cite{cfghpv} and references therein).
\item{2]} Historically it was the first to be studied in QCD. The so called 
``L\"uscher term'' actually is nothing but
the dominant contribution of
this bosonic string correction in the $2R<L$ limit.
\item{3]} It has been recently observed that it well describes the finite size
corrections of the interquark potential in SU(3) LGT  both extracted from
Wilson loops~\cite{ns01} and from Polyakov loops correlators~\cite{lw02} and
also in SU(N) LGT with $N\not=3$~\cite{lt01} (see also the analysis
of~\cite{gp99})). 
\end{description}
 Notwithstanding being the simplest one, this CFT 
is all the same highly non-trivial. In
 particular, as we shall see in detail in sect. \ref{s3}, the peculiar  choice of
 lattice sizes which is usually made in standard lattice simulations 
requires that one takes into account the {\sl whole} functional form of 
$F_q^1(R,L)$,
and not only the dominant contributions discussed in eq.s
(\ref{zbig}) and (\ref{zsmall}).
This is indeed
 one of the main points of this paper and we shall discuss it in detail in 
 sect. 4 when comparing our predictions with the numerical simulations. 

The whole functional form of  $F_q^1(R,L)$ can be evaluated by a suitable
 regularization of the Laplacian determinant (or alternatively by summing over
 the whole set of states of the Virasoro algebra).
This result has a rather long history: it was discussed for 
the first time in 1978 by M. Minami in~\cite{minami}. It
was then reobtained
in ref.s~\cite{df83,flensburg} and with a different approach 
in ref. \cite{fsst}. Here, we only report the
 result, which  for a $d$ dimensional gauge theory
 (i.e. $d-2$ bosonic fields) is:

\eq
F_q^1(R,L)=(d-2)\log\left({\eta(\tau)}\right)
\hskip0.5cm
;\hskip0.5cm {-i}\tau={L\over 2R}~~~,
\label{bos}
\en
\noindent
where $\eta$ denotes the Dedekind eta function:
\eq
\eta(\tau)=q^{1\over24}\prod_{n=1}^\infty(1-q^n)\hskip0.5cm
;\hskip0.5cmq=e^{2\pi i\tau}~~~,\label{eta}
\en
and $R$ is the distance between the two Polyakov loops.
\vskip.3cm

We list below for completeness the  
power expansions in the two 
regions: 

\begin{description}
\item{$2R<L$}
\eq
F_q^1(R,L)=\left[-\frac{\pi L}{24 R}
+\sum_{n=1}^\infty \log (1-e^{-\pi nL/R})\right](d-2)~~~,
\label{zsmalltot}
\en

\item{$2R>L$}
\eq
F_q^1(R,L)=\left[-\frac{\pi R}{6 L}+\frac{1}{2} \log\frac{2R}{L}
+\sum_{n=1}^\infty \log (1-e^{-4\pi nR/L})\right](d-2)~~~.
\label{zbigtot}
\en
\end{description}

These are the expressions that we shall compare in sect. 4 with our Montecarlo
data. Notice, as a side remark,  that for any practical purpose it is
enough to truncate the infinite sums which appear in eq.s~(\ref{zsmalltot}) 
and (\ref{zbigtot}) to the first two or three terms. The errors obtained in this way
(if one remains inside the regions of validity of the two expansions: 
$z<1$ for eq.(\ref{zsmalltot}) and
$z>1$ for eq.(\ref{zbigtot}))  are
much smaller than the uncertainties of the numerical estimates.

\subsection{The Nambu-Goto string}
\label{s2.5}
The approach discussed in the previous sections is very general. It shows that
at large enough interquark distance, the finite size corrections to the potential
are independent of the fine structure details of the effective
string model and only depend on the choice of boundary conditions, on the number
of transverse dimensions and on the geometry of the observable used to extract
the potential. This {\sl universality} of the string correction was already
observed by L\"uscher, Symanzik and 
Weisz in their original papers~\cite{lsw} and remains the nicest feature of the
effective string approach to the interquark potential. However, it is clear that
along this way we have no hope to predict the effect 
(or even simply check the
existence)
of  higher order corrections due to the self-interaction of the string.
These self-interaction terms are expected to play a role in the intermediate
region, before the asymptotic regime of the pure free bosonic CFT is reached. 
Notice however that there is no sharp separation between these
two regimes, and the border between them only depends on the resolution 
of the data used
to test the predictions. Precise enough data could allow to detect these higher
order terms (if they exist) at any value of the interquark distance.

In order to study the self-interaction of the string,
a precise choice of the effective string model (i.e. a precise form
for the world-sheet Lagrangian of the underlying string) is needed. In this paper we
shall follow the simplest possible option, which is known as the Nambu-Goto
action. There are a few reasons which support this choice:
\begin{itemize}
\item
It is the simplest and most natural 
generalization of the free bosonic CFT, since its action
is simply given by the area of the 
world--sheet, with no need of additional information or degree of freedom.
\item
It implies a behavior of the deconfinement temperature which seems to agree
rather well with the simulations (see sect. \ref{s2.6} below).
\item
As far as we know, there is only  one other case in
which higher order corrections to the free effective string  have been
detected and studied, i.e. the finite size behavior of the 
interface free energy in the three dimensional Ising model~\cite{cfghpv,pv94}. 
In this case the Montecarlo data were in perfect agreement
with the 
predictions obtained using the Nambu-Goto action.
\end{itemize}

However it is important to stress that this is by no means the only possible
choice. In fact there are several other actions which can give (at the first
order to which we are addressing the problem here) the same corrections.

Besides the self-interaction type terms, 
one could also include in the action ``boundary type'' terms, like
those studied in~\cite{lw02}. We decided in this paper to neglect this class of
higher order corrections, since they require the introduction of a free
parameter which must be fitted from the data and also because
here we are mainly
interested in the string self-interaction terms. However we plan to address
the issue of boundary correction in a forthcoming paper.

\subsubsection{Finite size corrections due to the Nambu-Goto string}
The major problem of the derivations that we shall discuss below is that
the gauge choice that we have to make in order to be able to perform our
calculations is not consistent at the quantum level. There are arguments which
tell us that this anomaly should vanish at large enough distance~\cite{olesen},
 but this
cannot eliminate the problem. This is the reason for which we repeatedly 
stressed in this paper that what we are addressing here is an {\sl effective}
string model. We are here in a completely different framework with respect to
the {\sl fundamental} string theories, for which consistence at the quantum
level is mandatory. 

We must think of the Nambu-Goto action as a low energy, large distance,
approximation of the ``true'' (unknown) fundamental string theory.

This can be clarified by looking at the 3d Ising model as an example. The
fundamental string is expected to describe the model at the microscopic level.
The 
common assumption is that it should describe the behavior (and the statistics)
 of the surfaces contained in the
strong coupling expansion of the model, as it does the free fermion field theory
in the 2d case. On the contrary, the effective string theory should describe the
behavior of these surfaces at a much larger distance scale, where the microscopic
features  become negligible and one only looks at the collective
modes of the fluctuations of these surfaces
which behave, as a first approximation, as free massless bosonic fields.

\vskip 0.3cm

As anticipated above, the Nambu-Goto string action is simply
given by the area of the
world--sheet:
\be
S=\sigma\int_0^{L}d\tau\int_0^{R} d\varsigma\sqrt{g}\ \ ,\label{action}
\ee
where $g$ is the determinant of the two--dimensional metric induced on
the world--sheet by the embedding in $R^d$:
\ba
g=\det(g_{\alpha\beta})&=&\det\ \de_\alpha X^\mu\de_\beta X^\mu\ \ .\\
&&(\alpha,\beta=\tau,\varsigma,\ \mu=1,\dots,d)\nonumber
\ea
 and $\sigma$ is the string tension.
\par
The reparametrization and Weyl invariances of the action (\ref{action})
require a gauge choice for quantization. We choose the ``physical gauge''
\ba
X^1&=&\tau\nonumber\\
X^2&=&\varsigma
\ea
so that $g$ is expressed as a function of the transverse degrees of freedom
only:
\ba
g&=&1+\de_\tau X^i\de_\tau X^i+\de_\varsigma X^i\de_\varsigma X^i\nonumber\\
&&\ \ \ +\de_\tau X^i\de_\tau X^i\de_\varsigma X^j\de_\varsigma X^j
-(\de_\tau X^i\de_\varsigma X^i)^2\\
&&\ \ \ \ (i=3,\dots,d)~.\nonumber
\label{use1}
\ea
The fields $X^i(\tau,\varsigma)$ must satisfy the boundary conditions
dictated by the problem. In our case periodic b.c.
 in one direction and Dirichlet b.c.
in the other one:
\be
X^i(0,\varsigma)=X^i(L,\varsigma); \hskip 1cm X^i(\tau,0)=X^i(\tau,R)=0\ \ .
\ee
It is clear that this  gauge fixing implicitly assumes that the surface is a
single valued function of $(\tau,\varsigma)$, i.e. it must not have overhangs or
cuts. This is certainly not the case for the microscopic surfaces which one
obtains in the strong coupling expansion. Thus this gauge fixing is just
another way to state
that the string that we are studying is an effective string. This point can be
made more rigorous by looking at the quantum consistency of this gauge fixing.
Indeed it is well known that
due to the Weyl anomaly this gauge choice can be performed at the
quantum level only in the
critical dimension $d=26$. However, in agreement with our picture of a large
scale effective string,
this anomaly is known to disappear at large distances \cite{olesen},
which is the region we are interested
in.\par

Inserting this result in eq.(\ref{action}) and setting for simplicity
$d=3$ (i.e. only one transverse degree of freedom)\footnote{With
this choice the quartic
terms in eq.(\ref{use1}) cancel out and the expression simplifies. Notice that
if one is interested in the action for $d>3$ these terms survive. This is the
reason for which in the final result one finds a non-trivial dependence on
$d$. A nice way to understand this fact is to notice
 that the self-interaction of the string also couples different transverse
 degrees of freedom.} we end up with
\eq
S[X]=\sigma\int_0^L \mbox{d}x_1\int_0^R \mbox{d}x_2
\sqrt{1+(\de_\tau X)^2+(\de_\varsigma X)^2} \;\;.
\en

Let us now expand the square root. As a first step,
in order to correctly identify the expansion parameter let
us  rewrite the action in terms of adimensional variables. Let us define:
$\phi=\sqrt{\sigma}X$,  $\xi_1=\tau/R$, $\xi_2=\varsigma/L$.
In this way we recognize that the expansion parameter is $(\sigma LR)^{{-1}}$.
Expanding the action keeping only the first two orders
(i.e. keeping only terms up to the fourth order in the fields) we find:
\eq
S[X]=\sigma L R+S^{\prime}(\phi) \;\;,  
\en
where
\eq
S^{\prime}(\phi)=S_{G}(\phi)-\frac{1}{8\sigma LR}S_{p}(\phi)+O\left((\sigma
LR)^{{-2}}\right) \;\;. 
\en

Let us look at these two terms in more detail:
\begin{itemize}
\item
 $S_{G}$ is a purely Gaussian term
\eq
S_{G}(\phi)=\frac12\int_{0}^{1}\mbox{d}\xi_{1}\int_{0}^{1}\mbox{d}\xi_{2}
\left(\nabla \phi\right)^2
\en
with
\eq
\left(\nabla \phi\right)^2=
\frac{1}{2u}\left(\frac{\de \phi}{\de \xi_{1}}\right)^{2}+2u
\left(\frac{\de \phi}{\de \xi_{2}}\right)^{2}
\en
and
\eq
u=\frac{L}{2R} \;\;. 
\en

It is easy to see that this term
is exactly the free
field action discussed in sect. \ref{s2.4}.
At this level of approximation the partition function becomes
\eq
Z(L,R)=\exp(-\sigma L R)\ Z_{1}\  \;\;, 
\en
where $Z_{1}$ is the Gaussian integral evaluated in sect. \ref{s2.4}
\eq
Z_{1}=\frac{1}{\eta(iu)} \;\;. 
\en
It is easy to see that this result also holds for $d>3$, each transverse
degree of freedom being
independent from the other so that the final result is simply the product of
$(d-2)$ times the Dedekind function. Thus we exactly recover the result of
eq.(\ref{bos}).

\item 
$S_{p}$ is the ``self-interaction term'':
\eq
S_{p}(\phi)=\int_{0}^{1}\mbox{d}\xi_{1}\int_{0}^{1}\mbox{d}\xi_{2}  
\left[\left(\nabla \phi\right)^2\right]^{2} \;\;. 
\en
At order $(\sigma LR)^{{-1}}$ the partition function is therefore 
\eq
Z(L,R)=\exp(-\sigma L R)\ Z_{1}\ \left(1+\frac{1}{8\sigma LR}\langle S_{p}
\rangle \right) \;\;, 
\en
where  the expectation value of $S_{p}$ is taken with respect to the 
action $S_{G}$.
Also  this expectation value can be evaluated using the
 $\zeta$-function regularization. The 
calculation can be found in~\cite{df83}:

\eq
\langle S_{p}\rangle=\frac{\pi^{2}}{36} 
u^{2}\left[2E_{4}(iu)-E_{2}^{2}(iu)\right] \;\;, 
\en

where $E_2$ and $E_4$ are the Eisenstein functions. The latter can be
expressed in power series:
\eqa
E_2(\tau)&=&1-24\sum_{n=1}^\infty \sigma(n) q^n\\
E_4(\tau)&=&1+240\sum_{n=1}^\infty \sigma_3(n) q^n\\
q&\equiv& e^{2\pi i\tau} \;\;, 
\ena
where $\sigma(n)$ and $\sigma_3(n)$ are, respectively, the sum of all
divisors of $n$ (including 1 and $n$), and the sum of their cubes.

\end{itemize}
\par

Bringing together the two terms we finally find (recall that we have fixed
$d=3$):
\eq
F_q^{(NLO)}(R,L)=\left[\log\eta(\tau)-\frac{\pi^2 L}{1152\ \sigma
R^3}\left[2
E_4(\tau)-E_2^2(\tau)\right]\right]+O\left(\frac{1}{\left(\sigma L
R\right)^2}\right) \;\;. 
\label{nlo}
\en
This is the functional form of the finite size corrections 
which we shall compare with the results of our Montecarlo simulations
in sect. 4.
Notice that the inclusion of next-to-leading terms does not require the
introduction of any new free parameter, so that the predictive power
is the same as for the free string case.

\subsection{Implications for the deconfinement transition}
\label{s2.6}
One of the most interesting consequences of eq.s~(\ref{zbig},\ref{zbigtot})
is that in the large $R$ limit the quantum
fluctuations of the flux tube are proportional to $R$ and 
have the effect to decrease the string tension. This change is proportional to
$T^2$ and introduce a dependence on the finite temperature of the effective
string tension:
\eq
\sigma(T)=\sigma(0)-\frac{\pi T^2 (d-2)}{6} \;\;,
\label{ol1}
\en
where $T=1/L$ denotes the finite temperature
 and $\sigma(0)$ is the zero temperature
limit of the string tension (which is measured, for instance,
through Wilson loop expectation values).
This process eventually leads to the deconfinement transition and can be used
(see~\cite{olesen_dec,pa82}) to estimate the adimensional ratio $\sigma(0)/T_c^2$.
If we assume that the free
string picture holds for all temperatures up to $T_c$,
eq.(\ref{ol1}) would predict the value of the latter to be
$T_c=\sqrt{6\sigma_0/\pi}$, a prediction that turns out to be rather far
from the value obtained in Montecarlo simulations.

This is another reason which supports the existence of higher order terms in the
effective string action. We can easily extend eq.(\ref{ol1}) so as to keep into
account the next to leading order in the Nambu-Goto action expansion. To this end, 
the modular transformation properties of the Eisenstein functions
\eqa
E_2(\tau)&=&-\left(\frac{i}{\tau}\right)^2
E_2\left(-\frac1\tau\right) +\frac{6i}{\pi\tau}\\
E_4(\tau)&=&\left(\frac{i}{\tau}\right)^4
E_4\left(-\frac1\tau\right)
\ena
turn out to be very useful.

Performing a modular transformation so as to reach the large $R$ limit we find
\eqa
E_2\left(i\frac{L}{2R}\right)&=&-\frac{4R^2}{L^2}\
E_2\left(i\frac{2R}{L}\right) +\frac{12 R}{\pi L}\sim
-\frac{4R^2}{L^2} \nonumber \\
E_4\left(i\frac{L}{2R}\right)&=&\frac{16R^4}{L^4} E_4
\left(i\frac{2R}{L}\right)\sim \frac{16R^4}{L^4}
\label{abc2}
\ena
so that
\eq
-\frac{1}{8\sigma LR}\langle S_{p}\rangle
\sim -\frac{\pi^2 R}{72\sigma L^3}
\en
and finally
\eq
F(L,R)\sim \sigma L R\left(1-\frac{\pi}{6 L^2\sigma}-\frac{\pi^2}{72
\sigma^2 L^4}\right) \;\;.
\label{ol2}
\en

This result perfectly agrees with the conjecture reported 
in~\cite{pa82,olesen_dec} which states that
if the world sheet bordered by the two Polyakov loops is
described by  a Nambu-Goto type action
 then the string tension should vanish at the critical point
with a square root singularity:  $\sigma(T)\sim (T_c-T)^{\frac12}$. This
behavior is compatible with eq.(\ref{ol1}) only if we assume:
\eq
\sigma(T)=\sigma(0)\sqrt{1-\frac{T^2}{T^2_c}}
\label{ol3}
\en
with 
\eq
T^2_c=\frac{3\sigma(0)}{\pi}~~~~,
\label{use2}
\en
which turns out to be in much better agreement with the results of MC
simulations.

Inserting this value into eq.(\ref{ol2}) we find:
\eq
F(L,R)\sim \sigma \frac{R}{T}
\left(1-\frac12\left(\frac{T}{T_c}\right)^2
-\frac18\left(\frac{T}{T_c}\right)^4\right)
\en

which is exactly the expansion to the next to leading order of eq.(\ref{ol3}).

Even if the estimate of eq.~(\ref{use2}) predicts a value for
the ratio $T_c^2/\sigma(0)$ which is in
  good agreement with the existing Montecarlo estimates for SU(N) LGTs, it  
  should be
considered with great caution, since it predicts a critical index 1/2 for the
deconfinement transition which  disagrees both with Montecarlo results and with
the expectations of the Svetitsky-Yaffe conjecture. This means that assuming
a Nambu-Goto type action is probably too naive and/or 
that the regularization of
higher perturbative
orders introduces new terms in the large R limit.

However, notwithstanding this cautionary observation, the previous
discussion certainly
tells us
that the simple free bosonic theory cannot be the end of
the story and that higher order terms must necessarily be present to match with
the expected behavior near the deconfinement transition.

\subsection{Range of validity of the effective string picture}
\label{s2.7}
As mentioned in the previous sections, the effective string picture is expected
to hold at large enough distances (see in particular the comments in
sect. \ref{s2.2}). However one of the surprising features of the
recent Montecarlo results~\cite{ns01,lw02}
 is that the effective string picture seems indeed to
hold at remarkably small distances. In~\cite{ns01} (Wilson loop operators in
$d=4$ SU(3) LGT) the range of validity starts at $R_c\sim 0.4$ fm. (In the
following we shall denote with $R_c$ the minimum value of interquark distance at
which we expect the effective string picture, possibly with higher order
corrections, to hold). A similar
result is also reported in~\cite{lw02} (Polyakov  loop correlators in
$d=3$ and $d=4$ SU(3) LGT), with $0.4\simlt R_c \simlt 0.5$ fm.

As for the Ising model, we found looking at the Wilson loop expectation values
that~\cite{wloops}  $\sigma
R_c^2\sim 1.5$ (see fig. 2 of ref.~\cite{wloops}). In exactly 
the same range of values also the logarithmic increase of the flux tube starts
to hold, in agreement with the effective string predictions (see the comment at
page 408 of~\cite{width}). If (with abuse of language) we try to write these
scales in fermi units using the definition of the Sommer scale $r_0$ which is
given by
$\sigma r_0^2=1.65$, we see that also in the Ising case we find
$0.4\simlt R_c \simlt 0.5$ fm. 
Following~\cite{wloops} we shall assume that also in our
present analysis $R_c=\sqrt{\frac{1.5}{\sigma}}$~.\footnote{The presence of this
threshold of validity is the main reason why earlier studies in the 3d Ising
gauge model, probing shorter
physical distances, due to the smaller computational power available, could not
identify the free bosonic string as the correct model and actually suggested a
fermionic string model~\cite{cgvf93}. 
It is now clear that, at least in the range of values
that we studied in the present paper, such a picture is not supported
by the data.}

All these observations show that the scale $R_c$ is much smaller than what one
would naively expect and that it seems to show a remarkable degree of
universality. It would be very interesting to understand the reason of this
behavior.

When dealing with Polyakov loop correlators, a natural scale to measure
distances is the critical temperature $T_c$, which is related to the string
tension by~\cite{ch96} 
\eq
\frac{T_c}{\sqrt{\sigma}}=1.2216(24) \;\;.
\en
For $L_c=1/T_c$ we hence get $\sigma L_c^2\sim 0.67$. This implies that $L_c\sim
0.3$ fm and  $R_c\sim 1.5 L_c$. 

In view of the above discussion it is useless to look at correlators below the
scale $L_c$, since in that region the string picture certainly does not
hold.
At the same time it is interesting to explore the scales below $R_c$ in the
range $L_c<R<R_c$ to see if the value of $R_c$ is again confirmed and/or if
higher order effects can in part take into account the deviation from the free
string picture below $R_c$.

\section{Simulations}
\label{s3}
\subsection{3d gauge Ising model}
\label{s3.1}

In order to test our predictions we performed a set of simulations on the 
 3d $\ZZ_2$ gauge 
model, whose partition function can be obtained from the general expression in
eq.~(\ref{eq1}) by setting $U_l\equiv \sigma_l \in \{1,-1\}$. The resulting
partition function turns out to be
\eq
Z_{gauge}(\beta)=\sum_{\{\sigma_l=\pm1\}}\exp\left(-\beta S_{gauge}\right)
~.
\en
The action $S_{gauge}$ is a sum over all the plaquettes of  a cubic lattice,
\eq
S_{gauge}=-\sum_{\Box}\sigma_\Box~~~,~~~
\sigma_\Box=\sigma_{l_1}\sigma_{l_2}\sigma_{l_3}\sigma_{l_4}~~.
\en

 As in eq.~(\ref{eq1}), we choose the same coupling in the time-like and in the
two space-like directions of the cubic lattice.

This model is known to have a roughening transition
at $\beta_{r}=0.47542(1)$\cite{rough_point},
and a bulk (i.e. at zero temperature)
deconfinement transition  at $\beta_{c}=0.7614133(22)$~\cite{dec_point}.
We performed our Montecarlo simulations  at three different values of the
coupling constant $\beta$, all located in the rough phase and close
enough to the deconfinement point to be well within the scaling
region. We chose three values for which the deconfinement temperature (and hence
the critical distance $R_c$) was known with high precision so as to be able to
precisely fix the minimal distance between the Polyakov loops and the lattice
size in the time direction.

It is important to recall that 
the 3d gauge Ising model can be translated into the  3d spin Ising 
model  by the so called Kramers-Wannier duality transformation 
\eqa
Z_{gauge}(\beta)&\propto& Z_{spin}(\tilde\beta)\\
\tilde{\beta}&=&-\um\log\left[\tanh(\beta)\right]~~,
\ena
where $Z_{spin}$ is the partition function of the Ising model in the 
dual lattice:
\eq
Z_{spin}({\tilde\beta})=\sum_{s_i=\pm1}\exp(-\tilde\beta H_1(s))
\en
with
\eq
H_1(s)=-\sum_{\Br ij \kt}J_{\Br ij \kt}s_is_j \;\;,
\en
where the sum runs 
over the links ${\Br ij \kt}$ connecting  the nearest-neighbor 
sites $i$ and $j$. Here the couplings $J_{\Br ij \kt}$ are 
fixed to the value $+1$ 
for all the links. This relation defines a one-to-one mapping between the
free energy densities in the thermodynamic limit. 

The expectation values of gauge invariant observables can be expressed 
as ratios of partition functions of the spin model. 
For instance the dual of
the Polyakov loop correlators, in which we are presently interested, is 
given by
\eq
\langle P(x)P^\dagger(x+R) \rangle =
\frac{Z_{spin,S}({\tilde\beta})}{Z_{spin}({\tilde\beta})} \;\;,
\en
where in $Z_{spin,S}$
all the couplings of the links (in the dual lattice) 
that intersect
a surface $S$ joining the two loops (any choice of the surface joining the two
loops is equivalent) take the value $J_{\Br ij \kt}=-1$. This
construction explains why the results that we are discussing 
are related (apart from the different choice of boundary conditions) to those 
obtained studying the interfaces of the 3d Ising spin model.

For the ratios of correlators that we shall study below, we get
\eq
\label{rpf}
\frac{G(R)}{G(R+1)} = \frac{Z_{spin,L \times R}({\tilde\beta})}
                           {Z_{spin,L \times (R+1) }({\tilde\beta})} \;\;,
\en
where we have taken the minimal surfaces that join the Polyakov loops.
Eq.~(\ref{rpf}) is the basis of the algorithm that we shall discuss below.

\subsection{The algorithm}
Computing the Polyakov-loop correlation function in the lattice gauge
theory in the straight forward way, the statistical error is increasing 
exponentially with $L$ and $R$. On the other hand the value
of $G(R)$ is decreasing exponentially with $R$. This problem is partially
resolved by the algorithm of L\"uscher and Weisz \cite{lw01}.

Our approach in the dual model overcomes the problem completely. 
The statistical error of the ratio $G(R)/G(R+1)$  virtually does not 
depend on $L$ and $R$.  The numerical results show that already for 
$R=2$ our method in the dual model gives similar 
statistical errors as the direct measurement in the gauge model.
For instance, in the $\beta=0.73107$ case, with $L=N_t=8$ we find with the present
algorithm $G(5)/G(4)=0.68421(9)$ (see tab. \ref{tab2a}) to be compared with the
value   $G(5)/G(4)=0.68429(21)$ obtained with the direct measurement and used in
our previous paper~\cite{cpp02}.

Our method is essentially an improved version of
 the so called ``snake algorithm'' introduced
in~\cite{fep00} to study the 't Hooft loop in SU(2) LGT's and later adapted to
the study of the interface free energy in the 3d spin Ising model~\cite{fp01}.
 The major
improvement in our algorithm with respect to ref.s~\cite{fep00,fp01}
is the hierarchical organization of the lattice updates (see below)
which allows us to greatly enhance the precision of our results.
Let us see in detail our algorithm.

\begin{figure}[htb]
\centerline{\epsfxsize=15truecm\epsffile{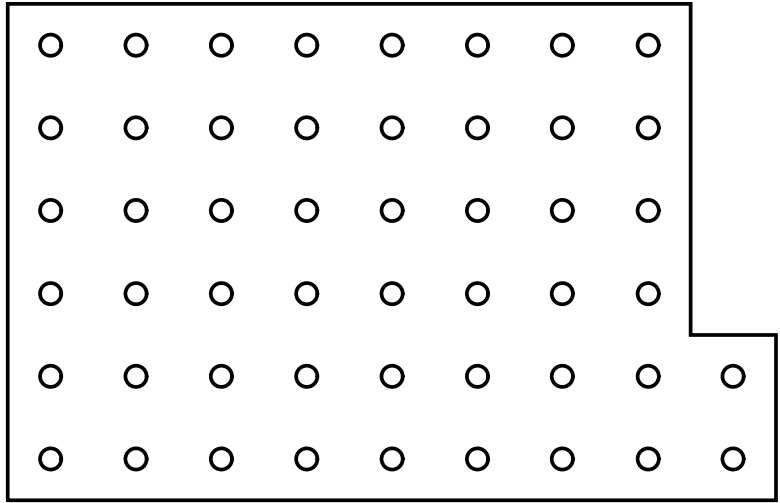}}
\caption{\label{surf}
Sketch of the surface denoted by $L \times R,M$. In the example, 
$L=6$, $R=8$ and $M=2$. The circles indicate the links  that intersect
the surface.
}
\label{sketch}
\end{figure}

In order to compute eq.~(\ref{rpf}) numerically, we factorize the ratio 
of partition functions in such a way that for each factor the partition
functions differ just by the value of $J_{\Br ij \kt}$ at a single link
\eq
\label{factorize}
\frac{Z_{L \times R}}{Z_{L \times (R+1)}}=
     \frac{Z_{L \times R,0}}{Z_{L \times R,1}} \; ... \;
     \frac{Z_{L \times R,M}}{Z_{L \times R,M+1}} \; ... \;
     \frac{Z_{L \times R,L-1}}{Z_{L \times R,L}} \;,
\en
where we have suppressed the index $spin$ and the argument $\tilde\beta$ to
simplify the notation. $L \times R,M$ denotes a surface that consists of 
a $L \times R$ rectangle with a $M \times 1$  column attached. 
A sketch is given in fig. \ref{surf}.

Each of the factors of eq.~(\ref{factorize}) can be written as expectation
value in one of the two ensembles:
\eq
\label{expectation}
\frac{Z_{L \times R,M+1}}{Z_{L \times R,M}} = 
\frac{\sum_{s_i=\pm1}\exp(-\tilde\beta H_{L \times R,M}(s)) \;
 \exp(- 2 \tilde\beta s_k s_l)}
     {Z_{L \times R,M}} \;\;,
\en
where $<k,l>$ is the link that is added going from
$L \times R,M$ to $L \times R,M+1$. Note that already the ratio of partition 
functions in eq.~(\ref{rpf}) could be written as an expectation value in the 
ensemble for the $L \times R$ surface. However the corresponding observable 
has an enormous variance.

The observable that we measure has only support on a single link on the 
lattice. Therefore it would be quite a waste of time to update the whole 
lattice before measuring $s_k s_l$.  In order to circumvent this problem 
we have enclosed the link $<kl>$ in a sequence of sub-lattices of the 
size $b_{1,i}  \times b_{2,i} \times b_{3,i}$. The center of each of the  
sub-lattices 
is the link $<kl>$ and $b_{1,i} \le b_{1,i+1}$, $b_{2,i} \le b_{2,i+1}$
and $b_{3,i} \le b_{3,i+1}$. $i$ is running from $1$ to $n$.  In our 
simulation we have taken $n=5$ throughout. 

Now we perform update sweeps over these boxes in a hierarchical way.
This can be best explained by the following piece of pseudo-code:
\begin{verbatim}
for(j_l=0;j_l<m_l;j_l++)
  {
  for(k_l=0;k_l<t_l;k_l++) sweep over the whole lattice;
  for(j_n=0;j_n<m_n;j_n++) 
    {
    for(k_n=0;k_n<t_n;k_n++) sweep over the sub-lattice n;
    for(j_nm1=0;j_nm1<m_nm1;j_nm1++) 
      { 
      for(k_nm1=0;k_nm1<t_nm1;k_nm1++) sweep over the sub-lattice n-1;
      .
      .
      . 
         for(j_1=0;j_1<m_1;j_1++)
            {
            for(i_1=0;i_1<t_1) sweep over the sub-lattice 1;
            measure s_k s_l;
            }
      .
      .
      .
      }
    }
  }
\end{verbatim}

As basic update algorithm we have used the microcanonical demon-update
with multi-spin coding implementation combined with a canonical update
of the demon \cite{Kari}.  Details on the implementation 
can be found in refs. \cite{hp93b,cfghpv}. In our implementation of 
multi-spin coding 32 or 64 lattices are simulated in parallel, depending 
on the architecture of the machine the program is running on. 
(Here we used Pentium 4 and Pentium III PC's; i.e. 32-bit machines. Hence
32 lattices are simulated in parallel).

In our simulations we have chosen the parameters of the algorithm ad hoc, 
without any attempt to optimize them. In particular we have always chosen 
5 sub-lattices of increasing size and $m_1=m_2=...=m_5=10$. 
As an example,  for $L=24$  we have chosen sub-lattice sizes of 
$2 \times 3 \times 3$, $4 \times 5 \times 5$, $8 \times 9 \times 9$,
$16 \times 17 \times 24$ and $32 \times 33 \times 24$. Note that the largest 
sub-lattices already take the full extent of the lattice in time 
direction.

In one cycle we 
performed $t_l=5$ for $\tilde \beta=0.228818$ and $\tilde \beta=0.236025$ 
and $t_l=10$ for $\tilde \beta=0.226102$
complete sweeps over the lattice. In one instance we did sweep
$t_1=t_2=...=t_5=2$ times over the sub-lattice.
In our production runs, we always 
performed $m_l=1000$ complete cycles. 
That means that the total number of measurements
for one value of $M$
is $32 \times 1000 \times 10^5 = 3.2 \times 10^{9}$.
Since $\exp(-\tilde \beta s_k s_l)$ can take only two values, the 
expectation value of $\exp(-\tilde \beta s_k s_l)$ can be easily obtained
from the expectation value of $s_k s_l$. Therefore, in the program, we 
accumulated $s_k s_l$ rather than $\exp(-\tilde \beta s_k s_l)$ itself. 
Averages of $s_k s_l$ over whole cycles were written to a file for later
analysis.

Before we started the measurement, 
8000 sweeps over the whole lattice
were performed for $\tilde \beta=0.228818$ 
and $\tilde \beta=0.236025$ and  16000 sweeps for 
$\tilde \beta=0.226102$  for equilibration.

In order to get a good estimate of autocorrelation times we performed one more
extended run for $\tilde \beta=0.226102$, $L=24$, $R=24$ and $M=0$
with 10000 complete cycles. 
For the cycle averages of $s_k s_l$ 
we obtain $\tau_{int} = 0.92(5)$ in units of cycles.
To deal with the large number of simulations, 
the analysis had to be automated.
Computing the statistical error, we performed a binning analysis with 
50 bins i.e. a bin size of 20 throughout. Given the small autocorrelation time,
this bin size should be sufficient.

As an example, we give the individual results for 
$Z_{L\times R,M+1}/Z_{L\times R,M}$ for $\tilde \beta=0.226102$, $L=24$ and
$R=24$ in table \ref{onelink}.
\begin{table}[h]
\begin{center}
\begin{tabular}{|r|c|}
\hline
 M & $Z_{L\times R,M+1}/Z_{L\times R,M}$ \\
\hline
 0 & 0.968311(25)  \\
 1 & 0.985483(29)  \\
 2 & 0.988214(23)  \\
 3 & 0.988921(23) \\ 
 4 & 0.989258(29) \\ 
 5 & 0.989387(24) \\
 6 & 0.989479(22) \\ 
 7 & 0.989538(28) \\
 8 & 0.989590(25) \\
 9 & 0.989560(24) \\
10 & 0.989665(26) \\
11 & 0.989660(27) \\
12 & 0.989720(27) \\
13 & 0.989742(30) \\
14 & 0.989750(28) \\
15 & 0.989775(28) \\
16 & 0.989780(26) \\
17 & 0.989866(26) \\
18 & 0.989988(32) \\
19 & 0.990072(29) \\
20 & 0.990396(29)  \\
21 & 0.991146(23) \\
22 & 0.993860(26)  \\
23 & 1.011208(30)  \\
\hline
\end{tabular}
\end{center}
\caption
{\sl  \label{onelink}
As an example we give the results for the ratios of partition functions defined
by eq.~(\ref{expectation})
$R=24$, $L=24$, $\tilde \beta=0.226102$ on a $128\times 128\times 24$ lattice.
The final result is $Z_{L\times (R+1)}/Z_{L\times R} = 0.77926(10)$.
}
\end{table}
Classically, one expects $Z_{L\times R,M}/Z_{L\times R,M+1}=\exp(-\sigma)=
\exp(-0.010560)=0.9895...$. In fact, the results for $1<M<22$ are rather 
close to this value. Note that $Z_{L\times R,0}/Z_{L\times R,1}$ is much
smaller and $Z_{L\times R,L-1}/Z_{L\times R,L}$ much larger 
than this value. I.e. it is unfavorable to create corners and favorable to 
eliminate them.

The statistical error of $Z_{L\times R+1}/Z_{L\times R}$ in table \ref{onelink}
is on average a little less than $0.00003$.
Given
the expectation value, it is easy to compute the variance
of $\exp(- 2 \tilde\beta s_k s_l)$, since
it can assume only the value $\exp( 2 \tilde\beta)$ or
$\exp(-2 \tilde\beta)$. Let us denote the probability  for the two signs
by $p_+$ and $p_-$. Then
\eq
p_+ + p_- =1
\en
and
\eq
p_+ \exp(+2 \tilde\beta)\; +\; p_- \exp(-2 \tilde\beta) =
\langle \exp(- 2 \tilde\beta s_k s_l) \rangle \approx 1 \;\; .
\en
Hence
\eq
 p_+ \approx \frac{1-\exp(-2 \tilde\beta)}
            {\exp(+2 \tilde\beta) - \exp(-2 \tilde\beta)}
            \;\;.
\en
The variance of $\exp(- 2 \tilde\beta s_k s_l)$ is hence given by
\eq
\mbox{var}[\exp(- 2 \tilde\beta s_k s_l)] \approx
 p_+ \exp(+4 \tilde\beta) + p_- \exp(-4 \tilde\beta) - 1 \;\;.
\en
For  $\tilde\beta=0.226102$ we get
$\mbox{var}[\exp(- 2 \tilde\beta s_k s_l)] \approx 0.208$. With this value for
the variance, the statistical error of $0.00003$ corresponds to
$N_{eff}  \approx 2.3 \times 10^8$ effectively independent
measurements, which does not compare too bad with the $3.2  \times 10^9$
measurements that actually have been performed.

Note that the results for the individual values of $M$ are obtained from
completely independent simulations, i.e.  computing the statistical error
of $Z_{L\times R}/Z_{L\times (R+1)}$ we can use standard error-propagation.
This becomes very simple, when we take the logarithm
\eq
 \log\left(\frac{Z_{L\times R}}{Z_{L\times (R+1)}}\right) = \sum_{M=0}^{L-1}
 \log\left(\frac{Z_{L\times R,M}}{Z_{L\times R,M+1}}\right) \;\;.
\en

Finally let us discuss the
CPU-time that was needed  for the simulations.
E.g. for the $128 \times 128 \times 24$ lattice at $\tilde \beta =0.226102$
the measurement for a single value of $M$ takes about $125$ min
on a P4 1.7 GHz PC. The time for equilibration is about $17$ min.
I.e. the total time to compute $G(R+1)/G(R)$ is $24\times (125+17)$ min
$=$ 2 days and 8 hours.

  \subsection{Comparison with other existing algorithms}
  The only other algorithm which allows an exponential reduction of the error in
  the
  measurement
  of large Polyakov loop correlators is the
  L\"uscher and Weisz algorithm (LW) discussed in~\cite{lw01} and ~\cite{lw02}.
  It is interesting to compare the performances of our algorithm with respect to
  LW. As mentioned above, the CPU time needed for the evaluation of a Polyakov
  loop correlator with our algorithm  increases  as $L$ ($L$ being the
  lattice size in the time-like
  direction)  and {\bf virtually does not depend}
  on the distance $R$ between the two loops\footnote{As a matter of fact, since
  with our approach each correlator
   requires an independent simulation, if one is interested
   in measuring
   the correlators for all the distances up to $R$ then
   the CPU time
   which is effectively required is linearly increasing
   with $R$.}. These two results (which are precisely confirmed by our
   simulations) both descend from the simple observation that,
   for a fixed number of iterations,
   the error of the ratio $\log Z_{L\times R,M}/Z_{L\times R,M+1}$ essentially
    does not depend on the arguments L,R and M.  Thus
   the statistical error of
   \eq
   \log(G(L,R)/G(L,R+1)) = \sum_{M=0}^{L-1} \log\frac{Z_{L\times R,M}}
   {Z_{L\times R,M+1}}
   \en
   has a negligible dependence on $R$,
   and increases as $\sqrt{L}$
   (because the number of terms that are added is
   $L$).
   Since the value of  $\log(G(L,R)/G(L,R+1))$ has again a negligible dependence
    on $R$ and is roughly proportional to $L$,
     we end up with the result that,
   if we keep
   the number of
   iterations fixed in each step of our iterative algorithm, then
   the relative error of $\log(G(L,R)/G(L,R+1))$  has a negligible
   dependence  on $R$ and decreases as $1/\sqrt{L}$.

   Since the numerical effort needed to keep this error dependence only
    increases as  $L^2$  (one power comes trivially from the change in
     lattice size, the  second from the fact that
      $L$ independent simulations are needed for each estimate of
      $\log(G(L,R)/G(L,R+1))$ ), the CPU time needed to keep
      a fixed relative  error increases as $L$.

    On the contrary, the CPU time, at fixed relative error, for the
   LW algorithm has a $L^3$ dependence on the length of the loops and
   {\bf exponentially increases}
   with $R$. Notwithstanding this exponential
   dependence, the LW algorithm represents a major improvement with respect
   to
   all other existing algorithms which would scale instead as $\exp(cRL)$.
   However,
   it is definitely less
   efficient than ours
   for large values of $R$. As
   a matter of fact, all the
   present studies with
   LW
   are confined to small or
   intermediate values of $R$ where only the short distance regime of the
   effective string can be studied,
   while our algorithm  allowed us for the
   first time to explore very large
   values of the interquark separation,
   where the large distance regime of the
   effective string can be  studied.

   It is also important to stress that since with our approach each correlator
   requires an independent simulation, the results
   have no cross-correlation. This
   is another major difference with respect to the LW algorithm, whose results
   are instead highly cross-correlated.

   In our opininion, in all the cases in which the dual transformation can
   be effectively implemented (hence in all the LGT's with abelian groups) a
   suitable generalization of our algorithm
   is always the best option to study
   Polyakov loop correlators,
    and also large Wilson loops.

\subsection{The simulation setting}
\label{s3.3}
We performed our simulations at three different values of $\beta$
and various choices of $N_t$, $N_s$ and $R$.
In choosing these values 
 we had to face four major constraints:
\begin{description}

\item{a]}
The lattice size in the space-like directions $N_s$ must be large enough so as to
avoid unwanted finite size effects due to the rather large values of the
correlation lengths that we shall study. 
Experience with the model suggests
that any value $N_s\geq 10\xi$ should solve this problem (see for
instance~\cite{nsgtxi}).

\item{b]} The distance $R$ between the two Polyakov loops should be larger than
the inverse of the critical temperature (see the discussion in sect. \ref{s2.7}) 
i.e. $R\geq L_c(\beta)= N_{t,c}(\beta)$. 

\item{c]} The values of $\beta$ that we choose must be in the scaling region.
This is needed at least for two reasons. The first  (and obvious) one
is that we do
not want to mix the finite size effects that we plan to observe (which are
expected to be very small) with unwanted effects due to scaling violations. The
second reason is that, as we shall discuss below, we need a very precise
estimate of the zero temperature string tension in order to perform our
analysis. 
This requires a careful extrapolation of the known values of the
string tension to the values of $\beta$ at which we
perform our simulations. 
Therefore high precision results for the string tension for a rather
dense set of $\beta$-values should be available in the literature.
Taking refs. \cite{hp93,cfghpv,hp97},
this means that we should have $\beta>0.73$.

\item{d]} We want to study the range of $T$ in which the precise functional form
of the string corrections is most important. This means that we should explore
the region $T_c>T>T_c/3$.

\end{description}

From the above discussion we see that a central role is played by the value of
$N_{t,c}$, so we decided to choose three values of $\beta$ for which the
critical temperature was known with high precision. The natural choice was
$\beta=0.73107$ for which $N_{t,c}=4$; 
$\beta=0.746035$ to which corresponds
  $N_{t,c}=6$
 and finally 
$\beta=0.75180$ for which $N_{t,c}=8$ (these values are taken from~\cite{ch96}).
 This choice fulfills constraint [c]. Further details on the parameters of the
 simulations can be found in tab. \ref{tab1},
 where we also list for completeness
 the values of the correlation length and
the zero temperature string tension for these three values of $\beta$. Since
the precise value of $\sigma(0,\beta)$ at the three values of $\beta$ that we
studied will play an important role in the following,
it is worthwhile to shortly discuss how we
extracted these estimates from the literature. The first important observation
is that, due to the high precision of our data and to the relative distance from
the critical point we cannot simply rely on the asymptotic scaling estimate of
$\sigma(0)$. A much better (and precise enough for our purpose)
estimate can be obtained taking as reference values those published
in~\cite{hp93,cfghpv,hp97}\footnote{We also used some unpublished values of
$\sigma$ obtained as a byproduct of the work published in~\cite{hp97}.}
and then interpolating among them with the law 
\eq
\sigma(0,\beta)=
\sigma(0,\beta_{ref})\left(\frac{\beta-\beta_c}{\beta_{ref}-\beta_c}\right)
^{2\nu}
\;\;, 
\en
where $\beta_{ref}$ is the coupling at which the reference value of $\sigma$ is
taken and $\beta_c$ is the critical temperature. If  $|\beta-\beta_{ref}|$ is
small enough, both the uncertainty in $\nu$ and the systematic error due to
neglecting higher order terms in the scaling law can be neglected. The
systematic error induced by this approximation can be estimated by repeating the
same analysis with another nearby value of $\beta_{ref}$. The
difference between the two results for $\sigma(0,\beta)$ obtained in this way
gives
a good estimate of this systematic error.
The final error on $\sigma$ is 
 the sum of the above  
systematic error plus the statistical errors
 of $\sigma(0,\beta_{ref})$ (as quoted in ref.s~\cite{hp93,cfghpv,hp97}). 

\begin{table}[h]
\begin{center}
\begin{tabular}{|c|c|c|r|l|l|r|}
\hline
\multicolumn{1}{|c}{$\beta$}
&\multicolumn{1}{|c}{$N_{t,c}$}
&\multicolumn{1}{|c}{$N_t$}
&\multicolumn{1}{|c}{$N_s$}
&\multicolumn{1}{|c}{$\xi$}
&\multicolumn{1}{|c}{$\sigma$}
&\multicolumn{1}{|c|}{$R_c$} \\
\hline
0.73107 & 4  &6,8,12& 64 & 1.41(3)&0.0440(3) & 5.84\\
\hline
0.74603 &  6 &9,12,18& 96 &2.09(4)&0.018943(32) & 8.90\\
\hline
0.75180 &  8  & 10,12,16,24 & 128 &2.95(10)&0.010560(18) & 11.92\\
\hline
\end{tabular}
\end{center}
\caption 
{\sl A few information on our simulations. In the first column the value of
$\beta$, in the second the inverse of the critical temperature. In the third and
fourth columns the
values of $N_t$ and $N_s$ that we studied. 
In the last three columns the values of the
correlation length, the zero temperature string tension and the corresponding
 value of $R_c$.}
\label{tab1}
\end{table}

Besides the simulations with the choice of parameters
 reported in tab. \ref{tab1} we also studied, in order to have a
cross-check of our finite size effect  predictions, (only in the case of
$\beta=0.74603$)  several  values of $N_t$ keeping $R$ fixed at the value
 $R=24$ (notice that for this value of $\beta$ we have $R_c\sim 9$ so with
 $R=24$ we are deep in the region of validity of the effective string picture).
The results are reported in tab. \ref{tab3}.

\section{Discussion of the results}
\label{s4}
For all the values of $\beta$ and $N_t$ listed in tab. 1 we extracted from the
simulations the expectation values of the ratios
\eq
\frac{G(R+1)}{G(R)}\equiv
\frac{\langle P(x)P^\dagger(x+R+1) \rangle}
{\langle P(x)P^\dagger(x+R) \rangle} \;\;.
\label{preq0}
\en 
We studied only a selected sample of values of $R$.
The results of the simulations are reported in tab.s~\ref{tab2a},\ref{tab2b} and
\ref{tab2c}.
Notice that, since each value of $R$ corresponds to a different simulation all
values reported in tab.s~\ref{tab2a}, \ref{tab2b} and
\ref{tab2c} are completely uncorrelated.

From these ratios we constructed the quantity:
\eq
Q_0(R)=\log\left(\frac{G(R)}{G(R+1)}\right)\;\;. 
\label{Q0}
\en
If no string effect is present, the correlator should follow the pure strong
coupling behavior of eq.(\ref{area}). 
Then it is easy to see that we should have
\eq
Q_{0,cl}=\sigma L \;\;. 
\en
So, in order to select the finite size corrections in which we are interested 
we defined:
\eq
Q_1(R)=\log\left(\frac{G(R)}{G(R+1)}\right)-\sigma L \;\;.
\label{Q1}
\en

We plot in fig.s 2-5 our data together with the prediction for $Q_1$ of the
pure string contribution (the simple Dedekind function) and the
Nambu-Goto correction eq.(\ref{nlo}) for the four values of $N_t$ at 
$\beta=0.75180$. For each value of $N_t$ we report in the figure caption the
value of $z_c\equiv \frac{2R_c}{N_t}$ beyond which the effective string picture
is expected to hold (see the discussion in sect. \ref{s2.7}).
The data for the other two values of $\beta$ show a similar
behavior as it can be easily checked using the values reported in
tab.s~\ref{tab2a},\ref{tab2b} and
\ref{tab2c}. 
\begin{table}[h]
\begin{center}
\begin{tabular}{|r|l|l|l|}
\hline
\multicolumn{1}{|c}{$R$}
&\multicolumn{1}{|c}{$N_t=6$}
&\multicolumn{1}{|c}{$N_t=8$}
&\multicolumn{1}{|c|}{$N_t=12$} \\
\hline
  4 &  0.77558(11) & 0.68421(9) &  0.55275(10) \\
  6 &  0.79682(14)&0.70519(10) &  0.57304(11)\\
  8 &  0.80850(14)& 0.71627(10) &  0.58320(13)\\
 10 &  0.81628(15)& 0.72351(11) &  0.58943(14)\\
 12 &  0.82151(18)& 0.72825(12) &  0.59387(16)\\
 14 &  0.82540(20)& 0.73183(14) &  0.59692(18)\\
 16 &  0.82798(20)& 0.73487(15) &  0.59956(19)\\
 20 &  0.83215(21)& 0.73870(16) &   0.60266(20)\\
 24 &  0.83564(25)& 0.74134(17) &  0.60522(22)\\
 28 &  0.83769(30)& 0.74332(21) &  0.60734(24)\\
\hline
\end{tabular}
\end{center}
\caption 
{\sl Values of the ratio of two successive Polyakov loop correlators:
$G(R+1)/G(R)$ for
various values of $R$ and $N_t$ at $\beta=0.73107$~.}
\label{tab2a}
\end{table}

\begin{table}[h]
\begin{center}
\begin{tabular}{|r|l|l|l|}
\hline
\multicolumn{1}{|c}{$R$}
&\multicolumn{1}{|c}{$N_t=9$}
&\multicolumn{1}{|c}{$N_t=12$}
&\multicolumn{1}{|c|}{$N_t=18$} \\
\hline
   6    &     0.84766(10) &  0.78156(7)& \\
   8   &      0.85912(11) &  0.79336(8)& \\
  10    &     0.86664(12) &  0.80081(9)& \\
  12    &     0.87219(13) &  0.80607(9)&  0.70586(7)\\
  14    &     0.87586(14) & 0.80992(11)&  0.70955(8)\\
  16    &     0.87908(15)& 0.81310(11)&  0.71235(8)\\
  18    &     0.88145(17) & 0.81539(13)&  0.71464(9\\
  20    &     0.88356(17) & 0.81731(13)&  0.71611(9)\\
  22    &     0.88468(17) &  0.81906(13)& \\
  24    &     0.88591(18) &  0.82037(13)& 0.71885(10)\\
  26    &     0.88725(20) &  0.82174(14)&\\
  28    &     0.88854(20) &  0.82249(15)& 0.72086(11)\\
  30    &     0.88941(20) &  0.82354(16)&\\
  32    &     0.89048(21) &  0.82407(15)& 0.72237(12)\\
  36    &&                  0.82570(17)&\\
  40    &&                  0.82691(18)&\\

\hline
\end{tabular}
\end{center}
\caption 
{\sl Same as tab. \ref{tab2a}, but with $\beta=0.74603$~.}
\label{tab2b}
\end{table}

\begin{table}[h]
\begin{center}
\begin{tabular}{|r|l|l|l|l|}
\hline
\multicolumn{1}{|c}{$R$}
&\multicolumn{1}{|c}{$N_t=10$}
&\multicolumn{1}{|c}{$N_t=12$}
&\multicolumn{1}{|c}{$N_t=16$}
&\multicolumn{1}{|c|}{$N_t=24$} \\
\hline
  8  &  0.91749(14)&0.88387(10) &0.83207(8)& 0.75075(7) \\
 12  &  0.92941(16) & 0.89646(12)&  0.84523(9)& 0.76492(8) \\
 16  &  0.93597(17) & 0.90351(14)& 0.85220(10)&0.77193(8) \\
 20  &  0.93999(18) &  0.90723(14)& 0.85655(12)& 0.77661(9) \\
 24  &  0.94292(19) & 0.91074(16)& 0.85937(13)& 0.77926(10) \\
 32  &  0.94706(21) &  0.91444(18)& 0.86371(16)& 0.78311(11) \\
 40  &  0.94864(20) &  0.91717(18) & 0.86624(15)& 0.78532(12) \\
 48  &  0.95111(23)& 0.91875(20)&0.86740(17)& 0.78677(13) \\
\hline
\end{tabular}
\end{center}
\caption 
{\sl Same as tab. \ref{tab2a}, but with $\beta=0.75180$~.}
\label{tab2c}
\end{table}

$Q_1(R)$ is affected by two different types of uncertainties. The one due to the
Polyakov loop correlators and that due to $\sigma$ (let us call it $\delta
\sigma$). The two must be treated
differently. We encoded the statistical errors of the Polyakov correlators as
usual with the error bars, while we kept into account the uncertainty in $\sigma$
by  plotting in the
figures (both for the Nambu Goto corrections and for the pure string term)
 two curves  obtained using $\sigma+\delta
\sigma$ and $\sigma-\delta\sigma$, respectively.

\begin{table}[h]
\begin{center}
\begin{tabular}{|r|c|c|c|c|}
\hline
 $N_t$&$z$&$Q_1\times 10^2$& $F^{NLO}\times 10^2$ & $F^{free}\times 10^2$ \\
\hline
 7 &  6.857  &     -7.63(3)& -7.70(2)&-5.439\\
  8 &  6.000 &     -6.15(3)& -6.05(3)&-4.504\\
 9&   5.333   &    -4.93(2)& -4.89(3)&-3.777\\
 10 &   4.800  &   -4.12(2)& -4.03(3)&-3.195\\
 11 &  4.364  &    -3.46(2)& -3.36(4)&-2.719\\
 12 &  4.000  &    -2.93(2)& -2.83(4)&-2.322\\
 14 &  3.428  &    -2.09(2)& -2.04(5)&-1.699\\
 16 & 3.000   &    -1.53(2)& -1.46(5)&-1.232\\
 18 & 2.667   &    -1.08(2)& -1.05(6)&-8.678\\
 24&  2.000   &    -0.31(2)& -0.24(8)&-0.140\\
\hline
\end{tabular}
\end{center}
\caption 
{\sl Results at $\beta=0.74603$, with $R=24$  kept fixed. 
 In the first column the value of
$N_t$ that we simulated. In the second column the corresponding values of
  $z$, in the third, the values of $Q_1$ obtained form the simulations. In the
  fourth column the prediction for $Q_1$ with the first correction due to the
  Nambu-Goto action. In the last column the corresponding quantity 
  obtained with
  the pure free bosonic string. These data are plotted in fig. 6.}
\label{tab3}
\end{table}

In fig. 6 we plot the data at fixed $R$ reported in tab. \ref{tab3}.

\begin{figure}[htb]
\centerline{\epsfxsize=15truecm\epsffile{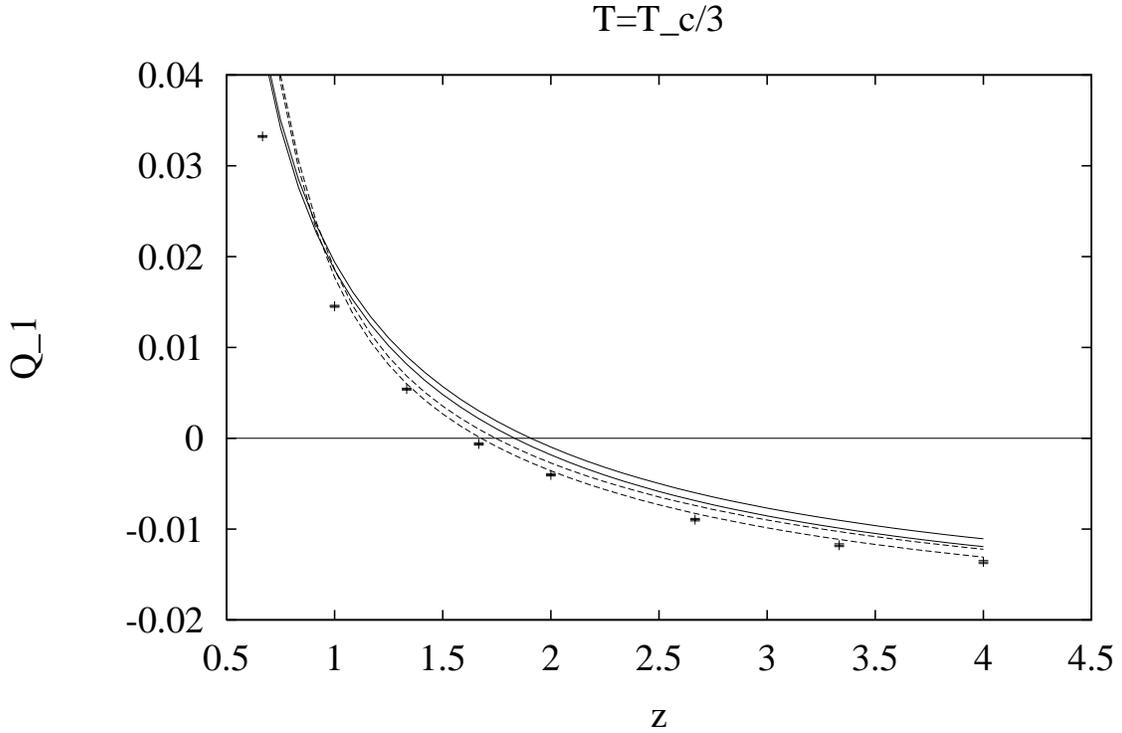}}
\caption{$Q_1$ for $N_t=24$ (i.e. $T=T_c/3$) at $\beta=0.75180$. The variable
$z$ is defined as $z\equiv \frac{2R}{N_t}$. The continuous
lines correspond to the free bosonic string prediction, 
while the two dashed lines
correspond to the first Nambu-Goto correction. The difference between the two
dashed and the two continuous 
lines keeps into account the uncertainty in our estimate of $\sigma$. The pure
area law corresponds to the line $Q_1=0$.
The threshold $z_c=2R_c/N_t$ beyond which the effective string picture is
expected to hold is located at $z_c\sim 1$ 
for these values of $N_t$ and $\beta$.}
\label{fig1}
\end{figure}
\begin{figure}[htb]
\centerline{\epsfxsize=15truecm\epsffile{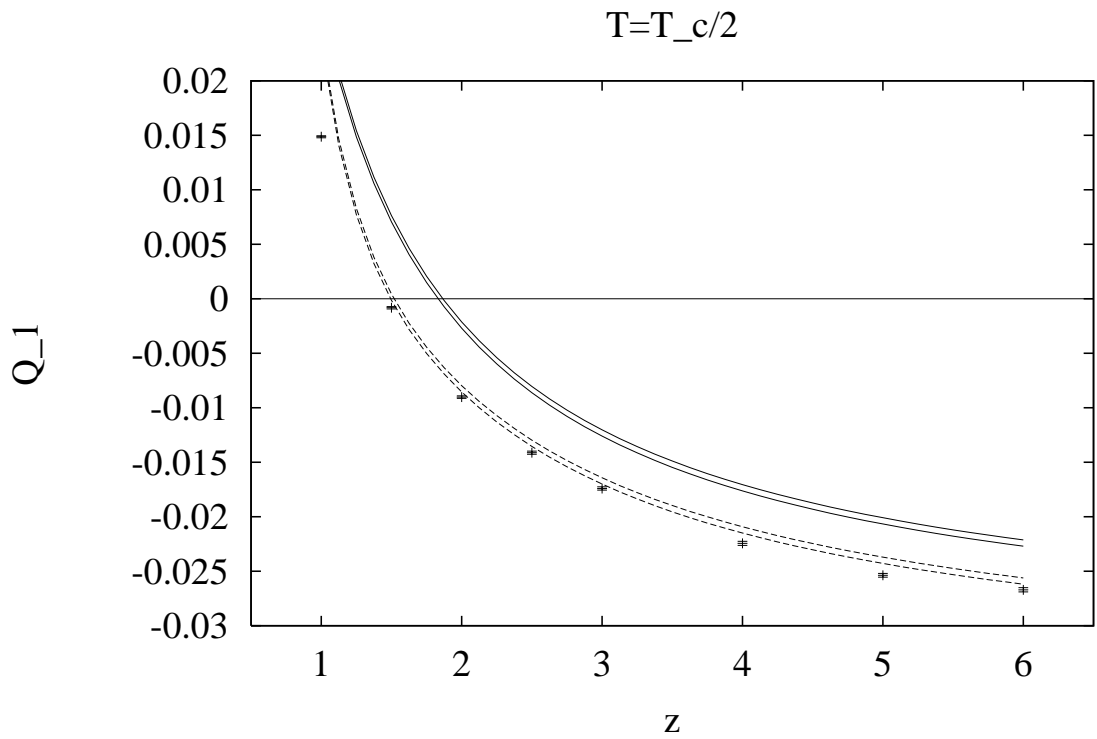}}
\caption{Same as fig. 2, but  for $N_t=16$ (i.e. $T=T_c/2$) at $\beta=0.75180$.
In this case we have $z_c\sim 1.5$.}
\label{fig2}
\end{figure}
\begin{figure}[htb]
\centerline{\epsfxsize=15truecm\epsffile{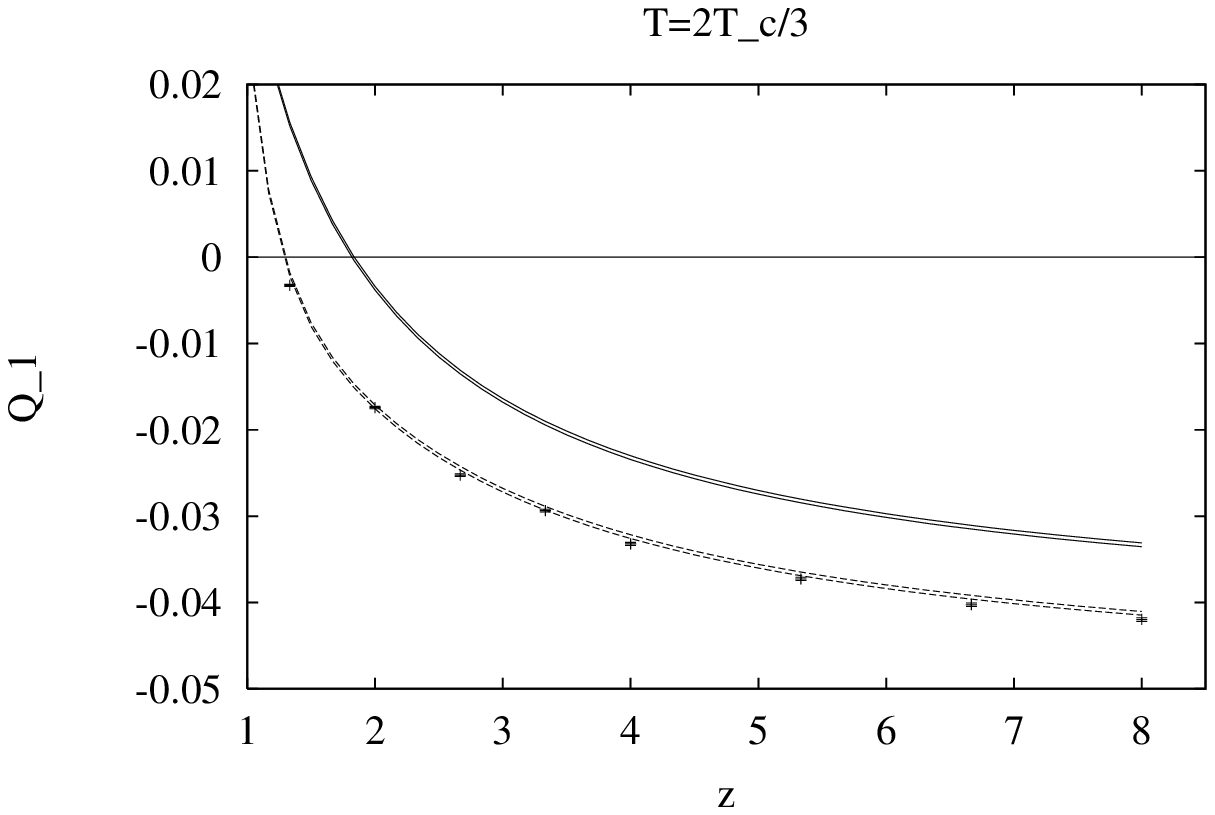}}
\caption{Same as fig. 2, but for $N_t=12$ (i.e. $T=2T_c/3$) at $\beta=0.75180$.
In this case we have $z_c\sim 2$.}
\label{fig3}
\end{figure}
\begin{figure}[htb]
\centerline{\epsfxsize=15truecm\epsffile{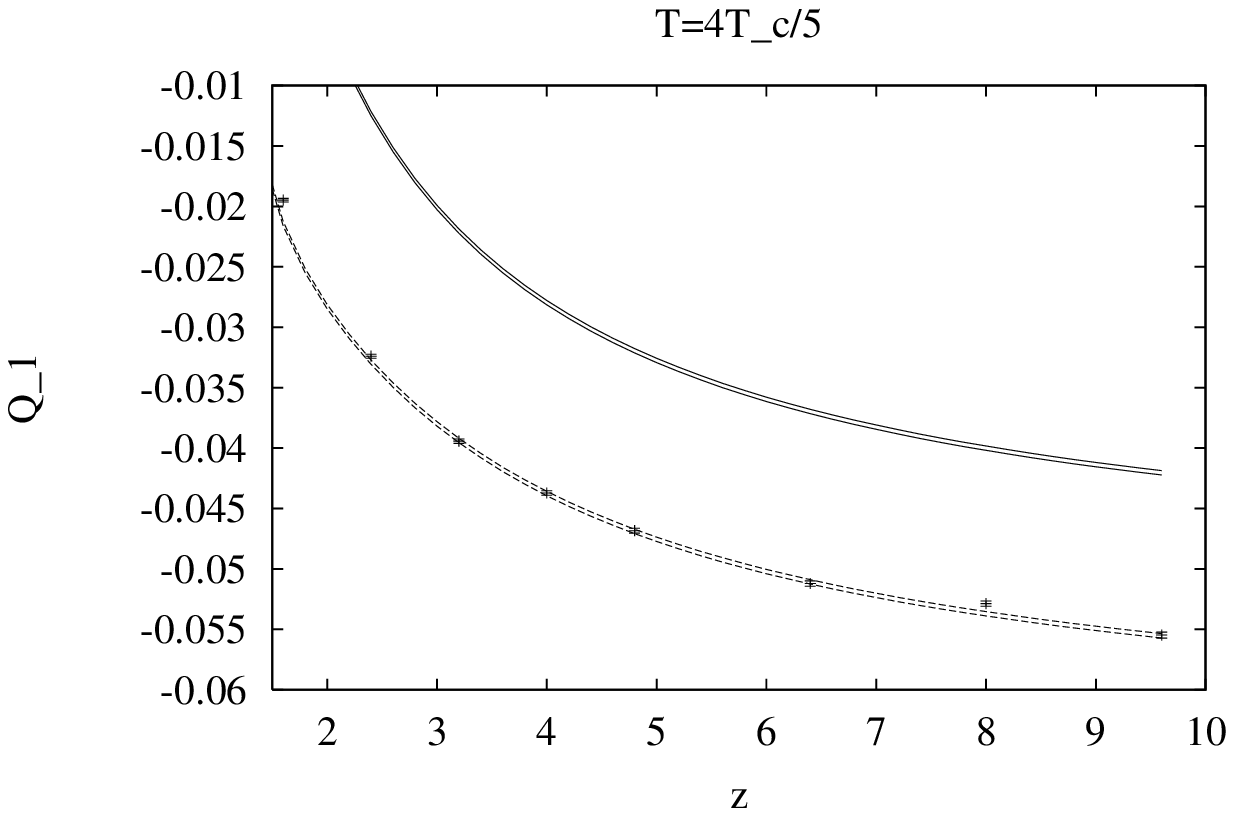}}
\caption{Same as fig. 2, but  for $N_t=10$ (i.e. $T=4T_c/5$) at $\beta=0.75180$.
In this case we have $z_c\sim 2.4$.}
\label{fig4}
\end{figure}
\begin{figure}[htb]
\centerline{\epsfxsize=15truecm\epsffile{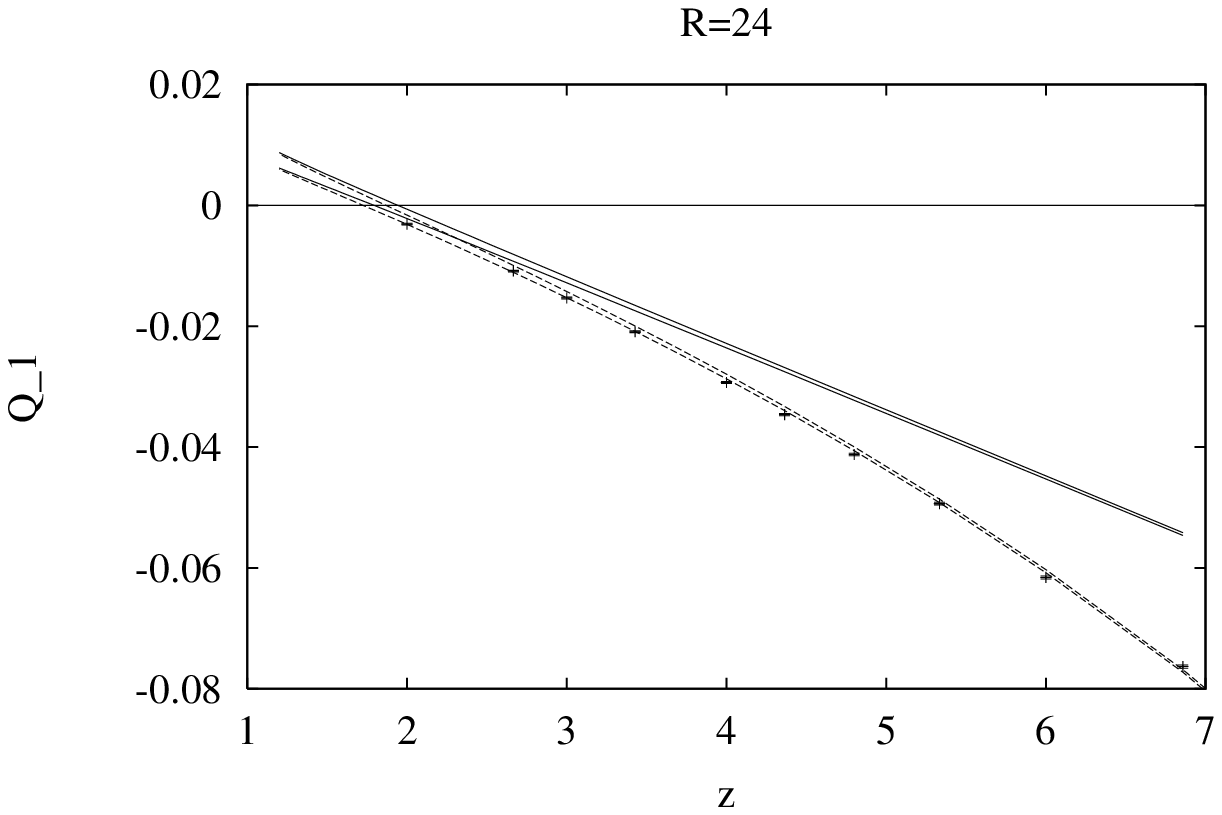}}
\caption{$Q_1$ for $R=24$ and the values of $N_t$ reported in tab. 3
 at $\beta=0.74603$. As in the
previous figures the continuous
lines correspond to the free string prediction while the two dashed lines
correspond to the first Nambu-Goto correction. For this value of $\beta$ we have
$R_c=9$.}
\label{fig5}
\end{figure}
Few comments are in order.
\begin{description}
\item{a]}
Finite size corrections with respect to the pure classical 
law of eq.(\ref{area}) are certainly present in the Polyakov loop correlators.
In fact the deviation from the pure area law  expectation (which is $Q_1=0$)
 is immediately
evident from the figures.

\item{b]} For the lowest temperature that we studied, i.e. $T=T_c/3$, the
contribution of the first correction of the Nambu-Goto
string is almost of the same order of magnitude of the uncertainty in the string
tension  (see fig. 2). 
This is the reason why we chose to study higher values of $T/T_c$
\footnote{Notice however that for very low values of $T/T_c$ (like those studied 
in~\cite{lw02}) the contribution due to the first order correction to the
Nambu-Goto string increases again in magnitude (but has the opposite sign) in
the small $R$ region (but has the opposite sign).}.

\item{c]} As the temperature increases, the gap between the pure bosonic string
prediction and the  Nambu-Goto one becomes larger and larger. It is clear,
looking at the figures, that the pure bosonic string  (continuous lines in 
fig.s 2-6) does not describe the data
in this temperature range.
The disagreement is most probably 
the signature of the fact
that the effective string underlying the 3d gauge Ising model is actually a
self-interacting string. It is easy to guess that the contribution due
 to this
 self-interaction becomes more and more important as the temperature increases,
 and this is indeed confirmed by the data.
 The data suggest that the main effect of this self-interaction is to lower the
 value of the string tension. As discussed in sect. 2.6,
 this effect is already present with the simple
 free bosonic string, but the lowering is enhanced by the self-interaction. 

\item{d]} There is a remarkable agreement between the data and the functional
form of eq.(\ref{nlo}) which describes the first correction of the Nambu-Goto
string with respect to the free bosonic string. It is important to stress that
this agreement is not the result of a fitting procedure. There is no free
parameter in eq. (\ref{nlo}). 

\item{e]} The agreement becomes worse and worse as $R$ decreases. The deviations
are particularly evident in the $R<R_c$ region ($z<z_c$ in fig.s 2-5).
This could be
simply due to the fact that one is approaching the size of the flux tube
thickness, but could also indicate that the functional form that we use is
inadequate  in the small $z$ region, or that a smooth cross-over is present
toward a different string picture.

\item{f]} The data show a very good scaling behavior: once the
proper value of $\sigma(\beta)$ is subtracted, no further dependence on $\beta$
is present (this is the reason why we needed very precise
independent estimates for $\sigma(\beta)$)

\item{g]} The agreement is particularly impressive for the last set of data,
i.e. those taken at fixed $R=24$ and shown in fig. 6. In this case, for
$N_t>12$ the 
prediction of eq.(\ref{nlo}) agrees with
the data within the errors (which, thanks to the nature of our algorithm, are
very small even if $R$ is large). It is worthwhile to notice that for this value
of $\beta$ we have $R_c\sim 9$.

\end{description}

In the definition of $Q_1$ we must insert the exact value of the string tension.
In principle it would be nice to avoid this external parameter and construct a
combination of Polyakov loop correlators in which only the effective string
corrections appear, without additional terms. This is easily achieved by the
combination
\eq
H(R,k)\equiv Q_1(R-k)-Q_1(R)
\label{defh}
\en
which is similar (apart from a different normalization) to the function $c(r)$
discussed in~\cite{lw02}.
However it is easy to see, looking at the large $R$ expansions of 
eq.(\ref{zbigtot}) and (\ref{abc2})
that this is not a good choice in the large $z$ region that we study here, where
the contributions to H(R) from the pure bosonic string and from the Nambu Goto
correction decrease as $1/R^2$ and $1/R^3$ respectively. On the contrary $H(R,k)$
 turns out to be a very useful quantity in the small $z$ region (as in the case 
 of~\cite{lw02}). This is well exemplified by fig. 7, where we plotted
 $H(R,1)$ as a function of $z$ for an hypothetical set of data with $L=60$ and $R$
 ranging from 6 to  30 (i.e. $z<1$). In fig. 8 we report $H(R,2)$ for our data at
 $\beta=0.74603$ and $L=12$. The data agree with the
 Nambu-Goto prediction, but the errors (even if very small) are of the same order
 of magnitude of the difference between the pure bosonic string and the Nambu-Goto
 correction. This is just another way to say that the major contribution of
 effective string fluctuations to the
 interquark potential in the large $z$ limit is simply a 
temperature dependence
of the string tension 
and it is exactly this signature that we observe looking at the
 $Q_1$ observable.

\begin{figure}[htb]
\centerline{\epsfxsize=15truecm\epsffile{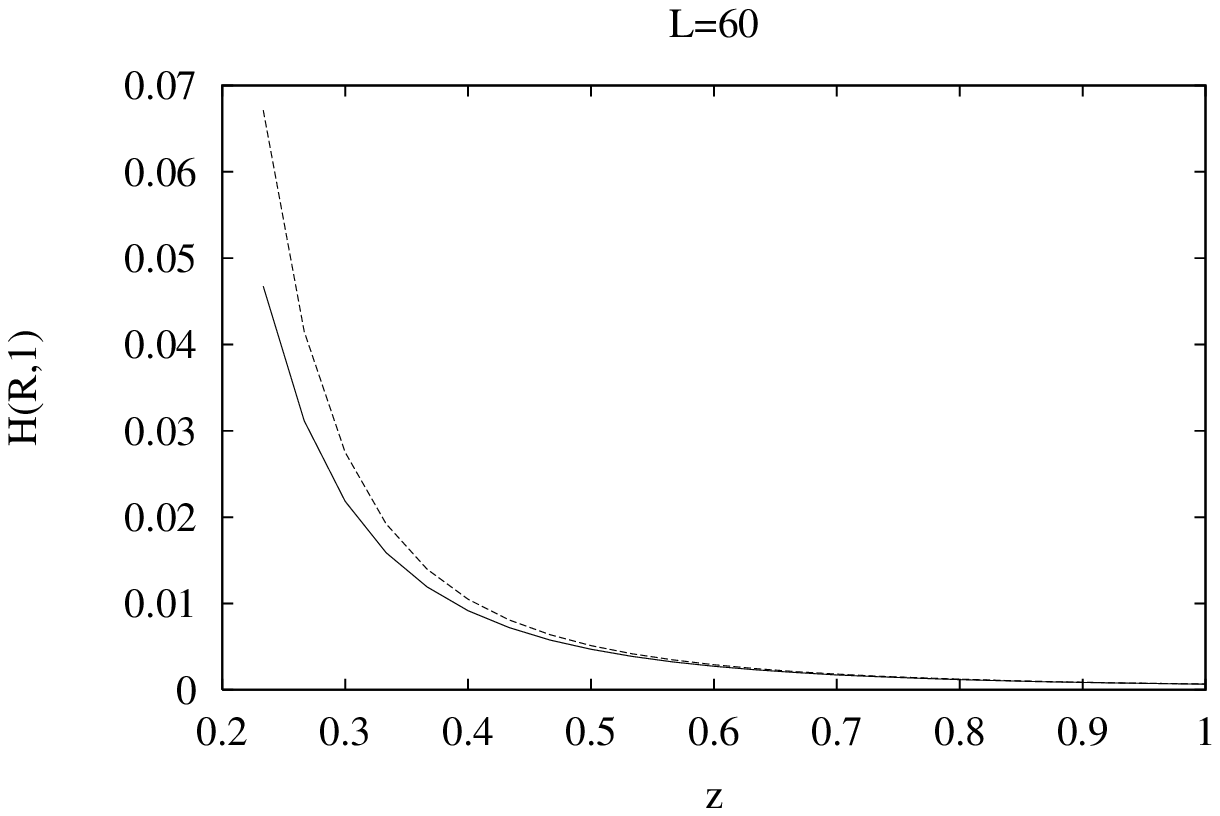}}
\caption{$H(R,1)$ as a function of $z$ for an hypothetical sample with $L=60$ and $R$
 ranging from 6 to  30 (i.e. $z<1$). The continuous line is the pure bosonic
 string correction, while the dashed line denotes the Nambu-Goto one.}
\label{fig7}
\end{figure}

\begin{figure}[htb]
\centerline{\epsfxsize=15truecm\epsffile{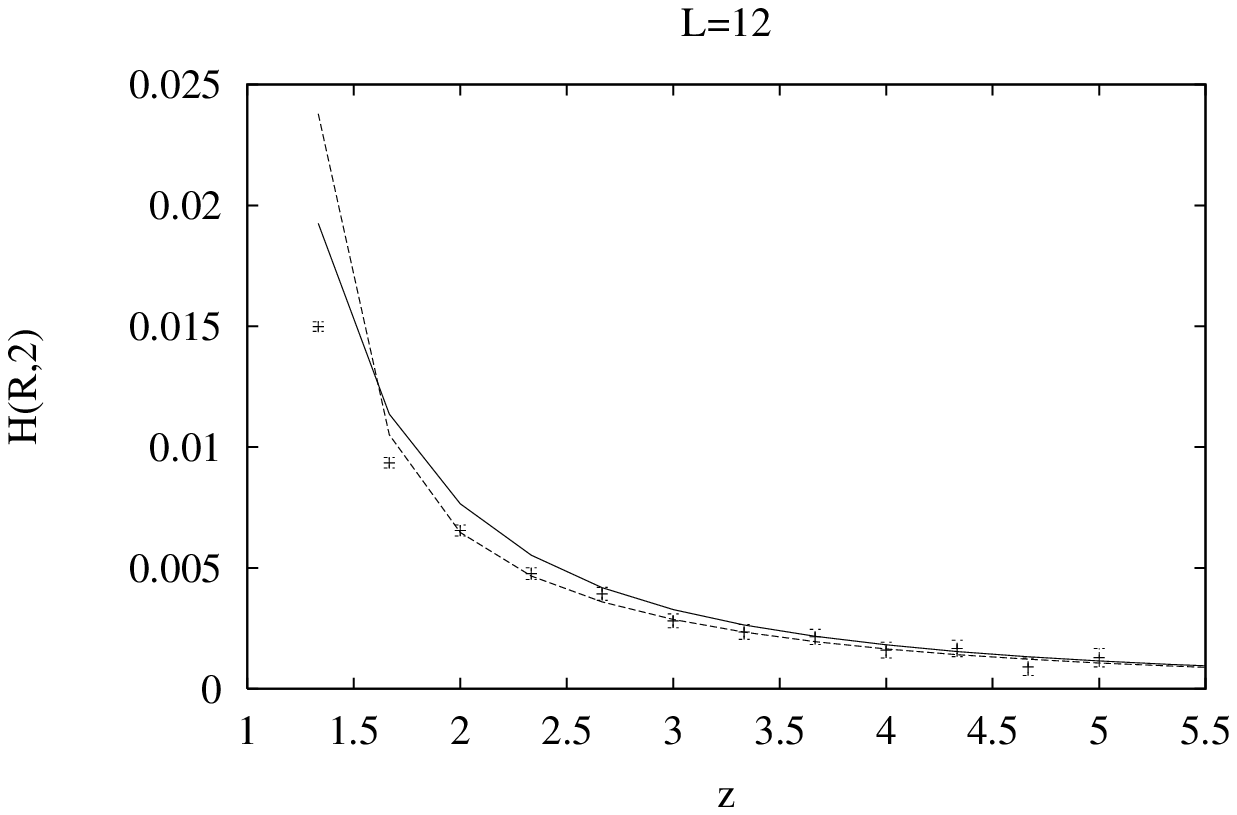}}
\caption{$H(R,2)$ for the data at
 $\beta=0.74603$ and $N_t=12$. The continuous line is the pure bosonic
 string correction, while the dashed line denotes the Nambu-Goto one. For this
 sample $z_c\sim 1.5$.}
\end{figure}

\section{Conclusions}
\label{s5}
Despite the impressive agreement which is manifest in fig.s 2-6, our analysis
leaves several open problems.  

\begin{description}
\item{1]} If we really assume that the Nambu-Goto proposal is the correct
description for the effective string, then the agreement that we find becomes
rather embarrassing, since there is apparently no room left for the higher order
corrections which one should expect in this framework. In principle one could
guess that these higher order correction are negligible, but this can be hardly
reconciled with the expected large $R$ behavior. For $R$ large enough, the
only term which survives is the finite renormalization of the string tension
which for
the Nambu-Goto string is expected~\cite{olesen_dec} to be simply given, order by
order, by the expansion of the square root in eq.(\ref{ol3}), and these terms
are certainly not negligible.

This observation agrees with the results recently appeared in~\cite{jmk02} where
the spectrum of string excitations in the $d=4$ SU(3) LGT was studied. The
observed spectrum seems to disagree both with the predictions of the pure bosonic
string and with that of the Nambu-Goto model. It would be interesting to study
the same spectrum directly in the 3d gauge Ising model.

\item{2]} Comparing our results with what L\"uscher and Weisz find in
the d=3 and 4  SU(3) LGT ~\cite{lw02} we see that the three models seem to be
described by three different effective string theories, with the same large
distance limit (the free bosonic string) but different
self-interaction terms. In fact in the d=3 case they find a perfect 
agreement with
the pure free string contribution and higher order self-interaction terms seem
to be absent. In d=4 they find higher order corrections which are modeled by a
boundary-type term. We checked that these deviations from the free string
behavior do not agree with the
$d=4$ version of the Nambu Goto correction of eq. (\ref{nlo}).
In principle there is no reason to
expect the same effective string in the 3d gauge Ising model and  in
the 3d and 4d SU(3) ones, however in past years 
it has become a common attitude to think that the
effective string model underlying a given LGT only feels the geometry of the
observables and is independent of the particular gauge group which one is
studying. The present numerical results and those of~\cite{lw02} 
suggest that this is not the case and that we are in presence of a variety of
different effective strings. In this respect it would be very interesting to
 have results from some
other models in $d=3$ so as to have a larger statistics and see if we are 
really  dealing with different effective strings~\cite{cpr02}.
 Notice, as a side remark, that
we are looking to a different range of values of $z$ with respect
to~\cite{lw02}. In principle it is also possible that the two LGT's show the same
behavior if they are studied in the same range of $z$ values.

\item{3]} Several independent results (and in particular the experience with the
dual problem of the interface fluctuations in the 3d Ising spin model)
suggest that in the 3d gauge 
Ising model the parameter which controls the effective string fluctuations is
the stiffness rather than the string tension. The two coincide at the critical
point, share the same leading scaling behavior in the scaling region, but have
different subleading corrections. Unfortunately there is presently no reliable
estimate of the stiffness in the scaling region, thus we are unable to estimate
the difference with respect to the string tension and evaluate the correction
that it induces in our estimates.  
However, since we study three different values of $\beta$, where this effect should
be quite different in amplitude, we are confident that our qualitative results
are not questioned by this problem.

\end{description}

Our results naturally raise the question whether the effective string underlying the
Ising model is of the Nambu-Goto type or not. In view of the above discussion, we
are not presently able to answer in a definite way. What we can state with
confidence is:
\begin{itemize}
\item
At large enough distances and low enough temperatures the data are well
described by a simple free effective bosonic string theory. Besides the Ising
models, the same seems to be true for SU(3) LGT in d=3 
and d=4~\cite{lw02,ns01}.

\item At shorter distances and/or higher temperatures, the effective string
picture still holds, but corrections due to boundary-type terms or to
self-interaction terms in the string action appear. The Montecarlo
 simulations suggest that these
corrections are different in the various models.

\item The peculiar geometry of the Polyakov loop correlators (in particular the
fact that the inverse temperature is related to the length of the Polyakov
loops) implies that they
are perfect tools to explore this region and detect higher order terms which
must necessarily 
show up, even for large values of $R$, as the critical temperature is
approached.

\item
At large distances these higher order terms
act to lower the string tension, while at short distances they behave
as $1/R^n$ corrections with $n>1$.

\item
In the 3d gauge Ising model that we studied in this paper the data remarkably
agree with the predictions of the Nambu-Goto action, truncated at the first
perturbative order.
\end{itemize}

It is clear from this discussion that there is still a very long way before we
can reach a precise
understanding of the effective string underlying lattice gauge theory.
 However
we are now in a much better position than before
 since new powerful numerical algorithms
(one of them is described in this paper) recently entered 
the game~\cite{lw01,lw02}.
The goal is worthwhile and certainly justifies further efforts in this
direction.

\vskip 1cm
{\bf  Acknowledgements}
We are deeply indebted to P. Provero for many useful suggestions and
discussions. We thank R. Sommer for a final reading of the draft.
This work was partially supported by the 
European Commission TMR programme HPRN-CT-2002-00325 (EUCLID).

\end{document}